\documentclass[secnumarabic,amsmath,amssymb,floatfix, nofootinbib,tightenlines,nobibnotes, aps, twocolumn]{revtex4}

\usepackage{xcolor} 
\usepackage[colorlinks=true,urlcolor=blue,linkcolor=blue, citecolor=blue]{hyperref}

\usepackage{graphicx}

\newcommand{\abs}[1]{\left| #1 \right|}
\newcommand{\bra}[1]{\left\langle #1 \right|}
\newcommand{\ket}[1]{\left| #1 \right\rangle}
\newcommand{\braket}[2]{\left\langle {#1{\left|\vphantom{#1 #2} \right.} #2} \right\rangle}

\newcommand{\sandwich}[3]{\left\langle {#1{\left| #2\vphantom{#1 #2 #3} |\right.} #3} \right\rangle}

\renewcommand{\epsilon}{\varepsilon}
\renewcommand{\phi}{\varphi}
\renewcommand{\Psi}{\varPsi}

\begin{document}

\begin{titlepage}
\title{Trajectory of a massive localised wave function in a curved spacetime geometry}
\author{Qasem Exirifard}
\email{qexirifa@uottawa.ca}

\author{Ebrahim Karimi}
\email{ekarimi@uottawa.ca}
\affiliation{Nexus for Quantum Technologies, University of Ottawa, 25 Templeton St., Ottawa, Ontario, K1N 6N5 Canada}
\begin{abstract}
Propagation of a localised wave function of a massive scalar field is investigated in its rest frame. The complete orthogonal Hermite-Gauss basis is presented, and the Gouy phase and Rayleigh scale notions are adapted. The leading and sub-leading gravitational corrections to a localised quantum wave function propagating in a generally curved spacetime geometry are calculated within the Fermi coordinates around the time-like geodesic of its rest frame, and cross-talk coefficients among the modes are derived. It is observed that spherically symmetric modes propagate along the geodesic. However, non-spherical modes are found to experience a mode-dependent residual quantum force at the sub-leading order. It is shown that the residual force does not generate an escape velocity for in-falling wave functions but leads to a mode-dependent deflection angle for the scattered ones.
\end{abstract}

\maketitle

\end{titlepage}

\section{Introduction}

The advancement of exchanging quantum information over long distances~\cite{Ursin:07,Yin:17,Yin:17PRL,2020npjQI,2020NJPh22i3074H} and the efforts to establish a quantum network in space~\cite{Dequal_2021,QuantumCommunication} have stemmed various studies to calculate the general relativistic corrections to the photon's wave function around the Earth or in a general background~\cite{Bruschi:2013sua,Bruschi:2014cma,Bruschi:2021all, Jonsson:2020npo, Barzel:2022tbf, DiPumpo:2022muv,Berera:2021xqa} and viability of quantum communication across interstellar distances is reported \cite{, Berera:2022nzs}. In particular, the authors of this manuscript utilize Fermi coordinates~\cite{Fermi:22} adapted to null geodesics~\cite{Blau:06} and report technologically measurable quantities beyond the light ray approximation~\cite{Exirifard:2020yuu,Exirifard:2021sfc}. Since these studies have considered the propagation of a wave function of massless particles along a null geodesic, one may ponder how a wave function of a massive particle propagates in a general curved spacetime geometry. The gravitational corrections for an electron bound to a  stationary or non-stationary hydrogen atom in a curved spacetime geometry have been investigated~\cite{Parker:1982nk,Parker:1980kw,Yu:2007wv,Zhou:2012eb,Cheng:2019tnk,Zhu:2008bd,Exirifard:2021cav}. However, the advancement of the structured quantum matter waves~\cite{Bliokh:2017uvr,RevModPhys.89.035004, 2018ConPh..59..126L} encourages examining the general relativistic corrections for a wave function of a freely propagating massive particle. We notice that the effects of constant Newtonian gravity are studied in~\cite{Hugo}. Here, we calculate the interaction between the Riemann tensor with the wave function of a localised massive field freely propagating along an arbitrary time-like geodesic in a general curved spacetime geometry. 

The paper is organised as follows: Section \ref{section:restframe} considers a localised wave function of a massive scalar field propagating in the flat Minkowski spacetime geometry. It generalises the concept of a rest frame of a particle to a rest frame of a wave function, and defines the rest frame of a wave function as the frame wherein the expectation value of the linear momentum vanishes. Section \ref{section:rfeq} studies the Klein-Gordon equation for a massive particle in the rest frame of the wave function. It considers the low energy modes and approximates the Klein-Gordon equation to the Schrodinger equation in $(3+1)$-dimensions.   Section \ref{section:generalhermit} reviews how to find a complete basis for the localized solutions of the \textcolor{black}{Schrodinger equation} in $(1+1)$-dimensions, and adapts the notion of the Gouy phase \cite{Gouy} and Rayleigh scale. Section \ref{section:elegantmodes} presents the elegant Hermite basis, and Section~\ref{Standard_basis} reviews the standard Hermite modes and their properties. Section \ref{Section:HermitModesin3D} presents the standard Hermite-Gaussian modes in $(3+1)$-dimensions as a complete orthogonal basis for the localised wave function of a massive particle.   

Section \ref{Section:curvedgeometry} considers a localised wave function of a massive particle propagating in a general curved spacetime geometry. It utilises the Fermi coordinates adapted to the time-like geodesic of the rest frame of the wave function up to the cubic order in terms of transverse directions \cite{Manasse:1963zz, doi:10.1063/1.524292}. The equation for the interaction between the curvature of the spacetime geometry is obtained. The interaction introduces a cross-talk between different modes and induces a distortion which is encoded in the operator $Q$ presented in \eqref{PsiQ}. The distortion operator at the quadratic and cubic orders are calculated in  \eqref{QOperator} and \eqref{Q3operator} and expressed in terms of the Riemann tensor and its covariant derivative evaluated and integrated over the time-like geodesic of the rest frame. Section \ref{Section:residualforce} utilises Newton's {\it Lex Secunda} to define the residual net force acting on the wave function $\Psi$ given by the time rate of the change of its average linear momentum. It reports that the residual net force vanishes at the level of quadratic corrections. Section \ref{secion:CubicCorrections} shows that the net residual force generally does not vanish at the order of the cubic corrections for a general wave function. The net residual force vanishes for spherical symmetric wave functions, so they follow the time-like geodesic of their rest frame. Non-spherically symmetric wave functions are found to experience a net residual force, and their mean trajectory deviates from the geodesic. 

Section \ref{section:schwarzchild} studies the Schwarzschild black hole \cite{SchwarzschildBHL} as an example of curved spacetime geometry. Section \ref{section:radialSch} calculates the net residual force for a localised wave function of a massive particle radially falling inside the black hole. It shows that though the force is not vanishing, as long as the width of the wave function on and outside the event horizon remains negligible compared to the Schwarzschild radius, the residual force does generate an escape velocity. Section \ref{section:radialSch} shows the effect of the residual force on the trajectory of a wave function deflected by the black hole. It reveals that different modes are deflected by an angle that depends on the mode number. Section \ref{section:conclusions} concludes that the computed dependency of the deflection of the mode number in the solar system is too small to be measured by available technologies. However, the finding that the trajectories of localised quantum wave functions deviate from geodesics may be counted as evidence supporting the idea that gravity should not be treated as a geometrical force in the quantum realm. 

\section{Rest frame of a wave function}
\label{section:restframe}
We consider $\phi(t,z^1,z^2,z^3)$ as a free complex-valued scalar field in the flat $(3+1)$-dimensional spacetime where the length of  line element in the spacetime is given by:
\begin{eqnarray}
	ds^2 = \eta_{\mu\nu} dz^\mu dz^\nu= - c^2 dt^2 + dz^i dz^i,
\end{eqnarray}
where $\mu, \nu \in \{0,1,2,3\}$, $i \in \{1,2,3\}$, $\eta_{\mu\nu}$ represents the components of the Minkowski metric, $c$ is the light speed, $dz^0=c t$,   $t$ represents time and $z^i$ represent space, and if a single index appears twice in a term, then summation is performed over that index.\textcolor{black}{}

The action for $\phi$ is given by:
\begin{eqnarray}
\label{RefereeEq2}
	S[\phi]=  -\int \textcolor{black}{ d^{4}z}\,\left(\eta^{\mu\nu} \partial_{z^\mu}\phi\, \partial_{z^\nu} \phi^* +\frac{\hbar^2 m^2}{c^2} \phi \phi^*\right),
\end{eqnarray}
where $\phi=\phi(z^i,t)$, $\phi^*$ is the complex conjugate of $\phi$, $\eta^{\mu\nu}$ is the inverse of the metric, $\partial_{z^\mu} = \frac{\partial}{\partial z^\mu}$, $d^4 z= dz^0 dz^1 dz^2 dz^3$, and $m$ is the mass while $\hbar$ represents the reduced Planck constant. Requiring that the functional variation of the action with respect to $\phi$ vanishes, generates the equation of motion of $\phi$:
\begin{eqnarray}
	\frac{\delta S[\phi]}{\delta \phi}=0 \to \left(\Box - \frac{\hbar^2 m^2}{c^2}\right) \phi(t,z^i) =0, 
\end{eqnarray}
where $\Box = -\partial_0^2 + \partial_i^2$. In natural units wherein $\hbar=c=1$, the equation of motion is simplified to:
\begin{eqnarray}
	(-\partial_t^2+\partial_{i}^2 - m^2 ) \phi(t,z^i) =0.
\end{eqnarray}
\textcolor{black}{While negative energy solutions and quantum field theory aspects can become relevant in curved spacetime, we emphasize that in the regime under consideration, the energy density and curvature of spacetime are sufficiently small that any potential effects of these solutions are expected to be negligible. Moreover, the specific physical processes we are considering do not involve significant particle production. Therefore, we only consider positive energy solutions to the equation of motion, which can be presented as follows}:
\begin{eqnarray}
\label{eq5}
    \phi(t,z^i)= \int \frac{d^3 p}{(2\pi)^{\frac{3}{2}}} A(\vec{p}) e^{i p_\mu z^\mu},
\end{eqnarray}
where $p_\mu=(E_{\!\vec{p}},\vec{p})$, $p_\mu z^\mu =- E_{\!\vec{p}} t + p_i z^i$, and 
\begin{eqnarray}
\label{Ep}
    E_{\!\vec{p}} = \sqrt{m^2 + p_1^2 + p_2^2 + p_3^2}.
\end{eqnarray}
We choose $\phi$ to be normalized, i.e.,
\begin{eqnarray}
    1= \int d^3 z \abs{\phi}^2 =\int d^3p \abs{A(\vec{p})}^2.
\end{eqnarray}
So $\abs{A(\vec{p})}^2$ can be interpreted as the probability for the particle to have momentum $\vec{p}$. We refer to $t, z^i$ as the coordinates in the lab. The average momentum in the lab is given by:
\begin{equation}
    {\vec{p}}_{_{\text{lab}}}= \langle \vec{p}\rangle_{\text{lab}}= \int d^3 p  \abs{A(\vec{p})}^2 \vec{p}.
\end{equation}
Without loss of generality, coordinates can be chosen such that momentum aligns with the positive direction of $z^3$:
\begin{equation}
    {\vec{p}}_{_{\text{lab}}}=(0,0,p_{\!_{\text{lab}}}),
\end{equation}
with $p_{\text{lab}}>0$. Let us consider a boost with velocity of $v=(v_1,v_2,v_3)$, which is given by the following Lorentz transformation:
\begin{equation}\Lambda^\mu_{~\nu}=
\left(
\begin{array}{cccc}
 \gamma  & -\gamma  v_1 & -\gamma  v_2 & -\gamma  v_3 \\
 -\gamma  v_1 & \frac{(\gamma -1) v_1^2}{v^2}+1 & \frac{(\gamma -1)v_1 v_2}{v^2} & \frac{(\gamma -1) v_1 v_3}{v^2} \\
 -\gamma  v_2 & \frac{(\gamma -1) v_1 v_2}{v^2} & \frac{(\gamma -1) v_2^2}{v^2}+1 & \frac{(\gamma -1) v_2 v_3}{v^2} \\
 -\gamma  v_3 & \frac{(\gamma -1) v_1 v_3}{v^2} & \frac{(\gamma -1) v_2 v_3}{v^2} & \frac{(\gamma -1) v_3^2}{v^2}+1 \\
\end{array}
\right),
\end{equation}
where $\gamma$ is the boost factor, $\gamma=\frac{1}{\sqrt{1-v^2}}$, and $v=\sqrt{v_1^2+v_2^2+v_3^2}$. We represent the new coordinates by $x^\mu=(\tau,x^1,x^2,x^3)$:
\begin{eqnarray}
    x^\mu =\Lambda^\mu_{~\nu} z^\nu.
\end{eqnarray}
The Fourier transformation of $\phi(\tau,x^i)$ is given by:
\begin{equation}
\label{RefEq12}
        \phi(\tau,x^i)= \int \frac{d^3 q}{(2\pi)^{\frac{3}{2}}} \tilde{A}(\vec{q}) e^{i q_\mu \textcolor{black}{x^\mu}},
\end{equation}
where $\vec{q}$ is the momentum in the $x$-frame and $\tilde{A}(\vec{q})$ is the amplitude for momentum $\vec{q}$. Since $\phi$ is scalar, its value in the $x$-frame is given by $\phi(\tau,x^i) = \phi(t,z^i)$ which implies:
\begin{equation}
    \tilde{A}(\vec{q}) = \frac{\partial(p_1, \cdots, p_D)}{\partial(q_1,\cdots, q_D)}A(\vec{p}),
\end{equation}
where $\frac{\partial(p_1, \cdots, p_D)}{\partial(q_1,\cdots, q_D)}$ is the Jacobian determinate. It can be simplified to:
\begin{equation}
   \frac{\partial(p_1, \cdots, p_D)}{\partial(q_1,\cdots, q_D)}=  \frac{E_{\!\vec{p}}}{\gamma(E_{\!\vec{p}} -p.v)},
\end{equation}
where $p.v=\sum_{i}p_i v_i$, $\gamma=\frac{1}{\sqrt{1-|v|^2}}$  is the boost factor where $\vec{v}$ is the velocity of frame $x$ with respect to the frame of $z$, and $E_{\!\vec{p}}$ is given in Eq.~\eqref{Ep}. The average of momentum in the $x$ frame is defined by:
\begin{equation}
    \langle q^\mu \rangle_x = \int d^3 q \abs{\tilde{A}(\vec{q})}^2 q^\mu,
\end{equation}
which is equal to:
\begin{equation}
 \label{qxEq}
   \langle q^\mu \rangle_x = \frac{\Lambda^\mu_{~\nu~}}{\gamma}\langle\frac{E_{\!\vec{p}}\, p^\nu}{E_{\!\vec{p}}- p.v}  \rangle_{\text{lab}}.
\end{equation}
Here, subscript x stands for the expectation value in the x frame.
We define the rest frame of a wave function as a frame wherein the expectation values of the linear momentum vanishes:
\begin{equation}
 \label{qxRest}
    \langle q^1 \rangle_{\text{rest}}=\langle q^2 \rangle_{\text{rest}}=\langle q^3 \rangle_{\text{rest}}=0,
\end{equation}
where the subscript of rest stands for the rest frame. Substituting Eq.~\eqref{qxEq} in Eq.~\eqref{qxRest} provides three equations that can be solved to find $\vec{v}=(v_1,v_2,v_3)$. In $(1+1)$ dimensions, we can prove that for any differentiable $A(\vec{p})$, there always exists one solution that satisfies $\abs{\vec{v}}< 1$.  We conjecture that for any  infinitely differentiable $A(\vec{p})$, Eq.~\eqref{qxRest} admits only one solution for $\vec{v}$ that satisfies $\abs{\vec{v}}< 1$. We call this unique solution as the rest frame of the wave function.  

As an example, let us consider a simple Gaussian profile for $A(p)$ with the variance of $\sigma$ around \textcolor{black}{$\langle \vec{p}\rangle_{\text{lab}} = (0,0,\bar{p})$}:
\begin{equation}
    A(\vec{p})= \frac{1}{\pi^{\frac{3}{4}} \sigma^{\frac{3}{2}}} \exp
    \left(-\frac{p_1^2+p_2^2 +(p_3-\bar{p})^2}{\sigma^2}\right).
\end{equation}
The symmetric distribution of the wave function in the $\textcolor{black}{z}^1$ and $\textcolor{black}{z}^2$ plane implies that the velocity of the rest frame holds:
\begin{eqnarray}
    v_1=v_2=0.
\end{eqnarray}
Then the equation for $v_3$ can be simplified to: \begin{equation}
    v_3 \left\langle\frac{m^2+p_1^2 +p_2^2}{\sqrt{m^2+p_1^2 +p_2^2+p_3^2}-p_3 v_3 }\right\rangle_{\text{lab}}=\bar{p}.
\end{equation}
This equation cannot be solved in an exact analytic form. However, when $\sigma$ is small, i.e., ($\sigma^2 \ll \bar{p}^2+m^2$), the perturbative solution can be found:
\begin{equation}
\label{v3restframe}
    v_3 = \frac{\bar{p}}{\sqrt{m^2+\bar{p}^2}}\left(1-\frac{(m^2+2\bar{p}^2)}{4(m^2+\bar{p}^2)^2}\sigma^2+\cdots\right).
\end{equation}
The above expression shows that the rest frame velocity not only depends on the average momentum but is a function of the internal structure of the wave function as it depends on $\sigma$. 

\section{Wave function in the rest frame}
\label{section:rfeq}
We would like to study the field's configuration representing a localised wave function. We choose the rest frame of the localised wave function. We show the coordinates in the rest frame by $(\tau,x^i)$, where $\tau$ stands for the proper time \textcolor{black}{in the rest frame}. The wave function satisfies the Klein-Gordon equation:
\begin{equation}
    (\Box - m^2) \phi(\tau,x^i) =0.
\end{equation}
Now let the field $\phi$ be expressed in terms of the complex-valued scalar field $\tilde{\phi}$, as:
\begin{eqnarray}
	\label{PhiField}
    \phi(x,\tau)= e^{-i m \tau} \tilde{\phi}(x,\tau).
\end{eqnarray}
The equation of motion for $\tilde{\phi}$ can be simplified to:
\begin{equation}
    -\partial_{\tau}^2 \tilde{\phi} +2 i m \partial_\tau \tilde{\phi} =- \partial_i^2 \tilde{\phi}.  
\end{equation}
 We restrict our study to low-energy field configurations wherein the second derivative of $\tau$ on left-hand side of the above equation can be neglected:
\begin{equation}
	\label{ParaxialApproximation}
    \textcolor{black}{2 i m \partial_\tau \tilde{\phi}} \approx - \partial_i^2 \tilde{\phi}.
\end{equation}
The equation of motion for $\tilde{\phi}$ in our study, therefore, can be approximated to:
\begin{equation}
    i  \partial_\tau \tilde{\phi} + \frac{1}{2m}\partial_i^2 \tilde{\phi}=0,
\end{equation}
which is the Schro\"dinger equation for a particle with mass $m$ in $(3+1)$-dimensions.

\subsection{Localised solutions in (1+1)-dimensions}
\label{section:generalhermit}
To find \textcolor{black}{a general localized solution for a wave function}, we first consider the Schro\"dinger equation for a particle with mass $m$ in $(1+1)$-dimensions of $\tau$ and $x$:
\begin{equation}
\label{Eq1}
    i  \partial_\tau \tilde{\phi} + \frac{1}{2m}\partial_x^2 \tilde{\phi}=0.
\end{equation}
\textcolor{black}{Let it be emphasized that $x$ is a variable defined only for the purpose of presenting the position in $1+1$ dimensions.} Equation \eqref{Eq1} is a linear homogeneous partial differential equation. We assume that boundary conditions are linear and homogeneous too. This allows us to utilise the technique of separation of variables to solve Eq.~\eqref{Eq1}.   The technique of separation of variables tries to find a complete basis for the on-shell field configurations. We look for solutions in the form of:
\begin{eqnarray}
\label{Eq2}
    \tilde{\phi}(x,\tau)= \phi(x f(\tau))~g(\tau),
\end{eqnarray}
\textcolor{black}{where $\phi$ and ($f$, $g$) are analytic}. Substituting Eq.~\eqref{Eq2} in Eq.~\eqref{Eq1} yields:
\begin{eqnarray}
    \label{phi_z_equation}
    \phi''(y) + \textcolor{black}{\textcolor{black}{\tilde{\alpha}}} \textcolor{black}{y} \phi'(y) + \textcolor{black}{\textcolor{black}{\tilde{\beta}}} \phi(y) = 0, 
\end{eqnarray}
where  $\phi'(y)\equiv \frac{d\phi}{dy}$,  $\phi''(y)\equiv \frac{d^2\phi}{(dy)^2}$, and 
\begin{eqnarray}
\label{z_in_term_x_f}
    y &\equiv & x f(\tau),\\
    \label{alpha_f_equation}
     \textcolor{black}{\tilde{\alpha}} &\equiv & \frac{2im f'}{f^3},\\
    \label{alpha_g_equation}
    \textcolor{black}{\tilde{\beta}} &\equiv & \frac{2i m g'}{f^2 g},
\end{eqnarray}
where $f'\equiv\frac{df}{d\tau}$ and $g'\equiv\frac{dg}{d\tau}$. \textcolor{black}{Notice that $\textcolor{black}{\tilde{\alpha}}$, $\textcolor{black}{\tilde{\beta}}$ and $y$ are variables defined within this section to concisely present the solution to \eqref{Eq1}.}
%
%
The technique of separation of variables demands that $\textcolor{black}{\tilde{\alpha}}$ and $\textcolor{black}{\tilde{\beta}}$ should be constant. For a constant $\textcolor{black}{\tilde{\alpha}}$, Eq. \eqref{alpha_f_equation} can be solved to find $f$, which is given by,
\begin{eqnarray}
\label{f_solved}
    f(\tau) = \pm \sqrt{\frac{m}{i \tau \textcolor{black}{\tilde{\alpha}} +  c_1}},
\end{eqnarray}
where $c_1$ is the constant of integration. Parity operator in $x$ ($x \to -x$) maps the positive sign of $f(\tau)$ to its negative value. So without loss of generality, we can set:
\begin{eqnarray}
\label{f_solved_2}
    f(\tau) = \sqrt{\frac{m}{i \tau \textcolor{black}{\tilde{\alpha}} +  c_1}}.
\end{eqnarray}
Equation \eqref{f_solved} can be subsitute\textcolor{black}{d} in Eq.~\eqref{alpha_g_equation}, and thus $g$  can be found as,
\begin{eqnarray}
    \label{g_alpha_beta}
    g(\tau)= (i \tau \textcolor{black}{\tilde{\alpha}} +  c_1)^{-\frac{\textcolor{black}{\tilde{\beta}}}{2\textcolor{black}{\tilde{\alpha}}}} c_2,
\end{eqnarray}
where $c_2$ is the constant of integration. For constant $\textcolor{black}{\tilde{\alpha}}$ and $\textcolor{black}{\tilde{\beta}}$, Eq.~\eqref{phi_z_equation} can be utilized to find $\phi(z)$:
\begin{eqnarray}
    e^{\frac{\textcolor{black}{\tilde{\alpha}} y^2}{2}} \phi(y) &=& 
     C_1 H_{\frac{\textcolor{black}{\tilde{\beta}} -\textcolor{black}{\tilde{\alpha}} }{\textcolor{black}{\tilde{\alpha}} }}\left(\sqrt{\frac{\textcolor{black}{\tilde{\alpha}}}{2}} y\right)\nonumber\\
    &+&C_2 ~ _1F_1\left(-\frac{\textcolor{black}{\tilde{\beta}} -\textcolor{black}{\tilde{\alpha}} }{2 \textcolor{black}{\tilde{\alpha}} };\frac{1}{2};\frac{\textcolor{black}{\tilde{\alpha}}  y^2}{2}\right),
\end{eqnarray}
where $C_1$ and $C_2$ are constants of integration, $H_{\frac{\textcolor{black}{\tilde{\beta}} -\textcolor{black}{\tilde{\alpha}} }{\textcolor{black}{\tilde{\alpha}} }}$ is the Hermite polynomial of degree $\frac{\textcolor{black}{\tilde{\beta}} -\textcolor{black}{\tilde{\alpha}} }{\textcolor{black}{\tilde{\alpha}} }$, and $ _1F_1$ is the Kummer's (confluent hypergeometric) function. Kummer's function admits an irregular singular point at $y = \infty$. We look for regular solutions at $y = \infty$, and thus, we set $C_2=0$, and obtain:
\begin{eqnarray}
\label{phi_z_Hermit_n}
    \phi(y) = C_1 e^{-\frac{\textcolor{black}{\tilde{\alpha}} y^2}{2}} 
      H_{\frac{\textcolor{black}{\tilde{\beta}} -\textcolor{black}{\tilde{\alpha}} }{\textcolor{black}{\tilde{\alpha}} }}\left(\sqrt{\frac{\textcolor{black}{\tilde{\alpha}}}{2}} y\right).
\end{eqnarray}
Notice that the technique of separation of variables aims to find a complete basis for all field configurations. \textcolor{black}{ The Hermite polynomials for non-negative integer values of $\frac{\textcolor{black}{\tilde{\beta}} -\textcolor{black}{\tilde{\alpha}} }{\textcolor{black}{\tilde{\alpha}} }$ in \eqref{phi_z_Hermit_n} leads to a complete basis  which in Optics is known as the Hermite Gaussian basis.} In order to generate this basis, we set:
\begin{eqnarray}
\label{beta_alpha_relation}
    \textcolor{black}{\tilde{\beta}} &=& (n+1)\textcolor{black}{\tilde{\alpha}},
\end{eqnarray}
where $n \in  \{0\} \cup \mathbb{N}$. The  finiteness of $\int_{-\infty}^{\infty} dx |\tilde{\phi}|^2$ demands $\textcolor{black}{\tilde{\alpha}}>0$.

\subsection{Elegant Hermite basis}
\label{section:elegantmodes}
It is convenient to define $w_0 = (\frac{2 c_1}{\alpha m})^{\frac{1}{2}}$, and consider that $w_0$ is a non-negative real number, and set
\begin{eqnarray}
\label{tauR}
    \tau_{\!_R} &=& \frac{1}{2} m w_0^2,\\
 \label{ComplexW}
   \bar{w} &=& w_0 \left(1+  \frac{i \tau}{\tau_{\!_R}}\right)^{\frac{1}{2}}.
\end{eqnarray}
Then, express $\tilde{\phi}(x,\tau)$ in terms of $w_0$, \textcolor{black}{$\tau_R$} and $\bar{w}$:
\begin{equation}
	\label{FieldPhiTilde}
    \tilde{\phi}_n(x,\tau) = {C}_n ~{\bar{w}^{-(n+1)}} H_n\left(\frac{x}{\bar{w}}\right) \exp\left(-\frac{x^2}{\bar{w}^2}\right), 
\end{equation}
where the index $n$ on the left-hand side shows that it is the wave function for mode $n$, the constants of $c_2 C_1$ are abbreviated to the constant $C_n$. Notice that we are using a compact notation, and indeed, $\bar{w}$ is complex and is a function of $\tau$. $w_0$ defines \textcolor{black}{the width of the wave function} at $\tau=0$, and $|\bar{w}|$ represents the width of wave function at $\tau$. In analogy with optics, we refer to $\tau_{\!_R}$ as the Rayleigh time.   \textcolor{black}{The width of the wave function doubles at $\tau=\tau_R$: $||\bar{w}(\tau_{\!_R})||=2 w_0$}.

We require that $\tilde{\phi}_{\textcolor{black}{n}}$ to be normalised:
\begin{equation}
    \int dx \tilde{\phi}_n \tilde{\phi}_n^* = 1,
\end{equation}
which is satisfied for: 
\begin{equation}
    C_n =\left( \frac{2\Gamma(2n)}{2^{n} \Gamma(n)} \sqrt{\frac{\pi}{2}} \right)^{-\frac{1}{2}} w_0^{n+\frac{1}{2}},
\end{equation}
where $\Gamma$ represents the Gamma function: $\Gamma(n)=(n-1)!$.  To summarise, the normalised mode $n$ is given by:
\begin{eqnarray}
    \label{Elegant_Representation}
    \braket{x}{\bar{n}}=\tilde{\phi}_n&=& \frac{1}{\sqrt{w_0}} \left( \frac{2\Gamma(2n)}{2^{n} \Gamma(n)} \sqrt{\frac{\pi}{2}} \right)^{-\frac{1}{2}} \nonumber\\
    &\times&{\left(\frac{w_0}{\bar{w}}\right)^{n+1}} H_n\left(\frac{x}{\bar{w}}\right) e^{-\frac{x^2}{\bar{w}^2}},
\end{eqnarray}
where the bar sign over $n$ is used in  $\braket{x}{\bar{n}}$ to highlight that \textcolor{black}{it represents the $n$th ``elegant mode''. In fact, Eq.~\eqref{Elegant_Representation} is known as the elegant basis of Hermite-Gaussian modes \cite{Siegman:73}}. 
\begin{figure*}[t]
    \begin{tabular}{c @{\hspace{0.5cm}} c @{\hspace{0.5cm}} c }
        \includegraphics[width=.30\linewidth]{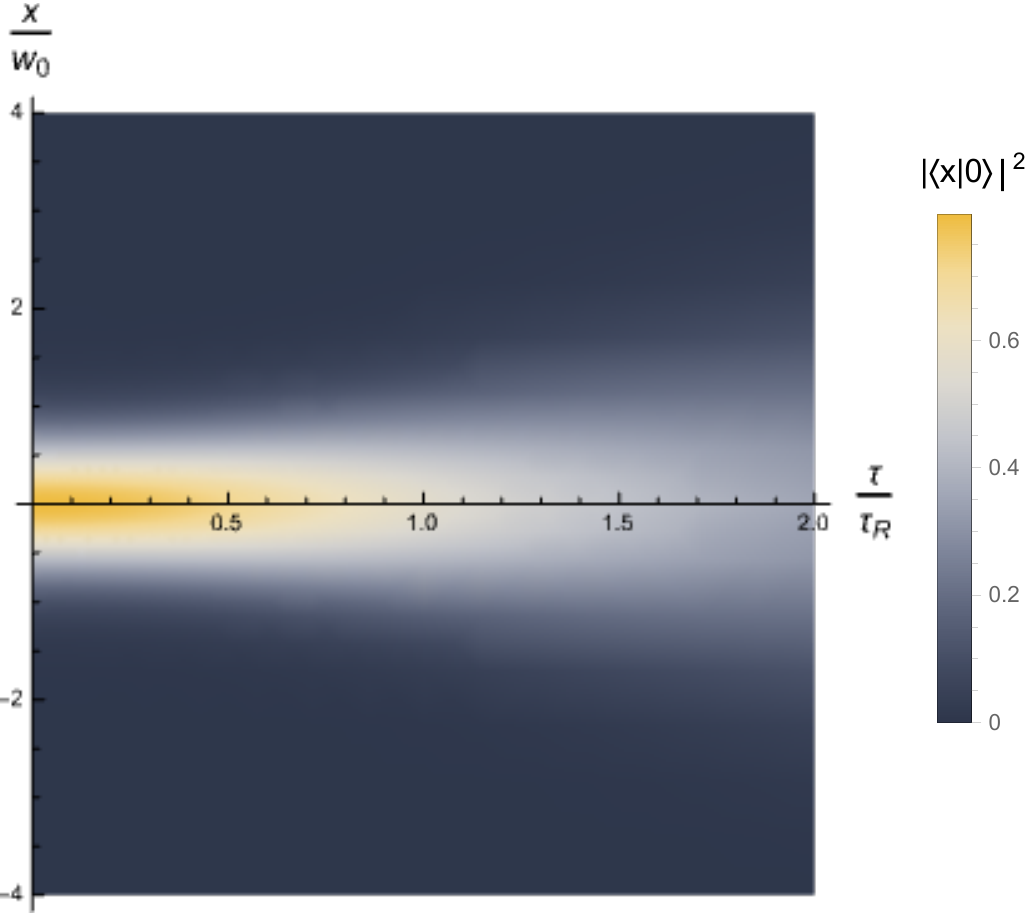}
        &
        \includegraphics[width=.30\linewidth]{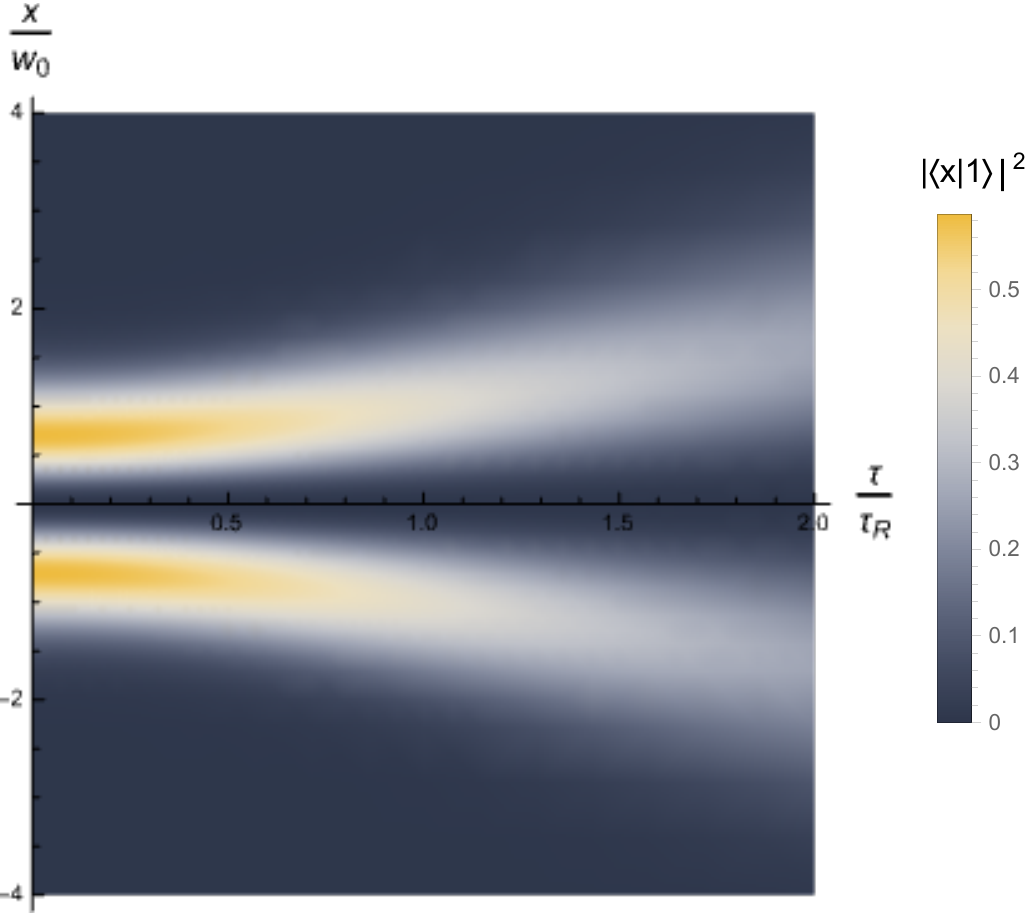}
        &
        \includegraphics[width=.30\linewidth]{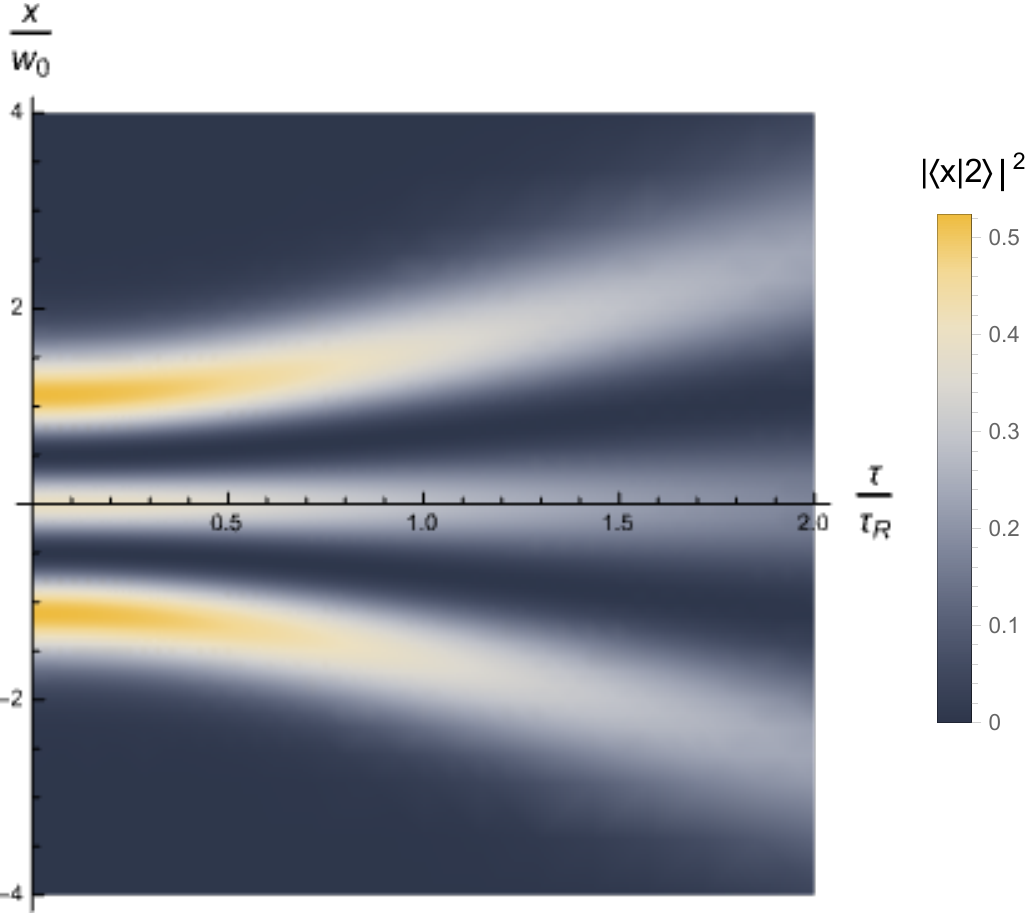}\\
        (a)  & (b)  & (c)  \\
        \includegraphics[width=.30\linewidth]{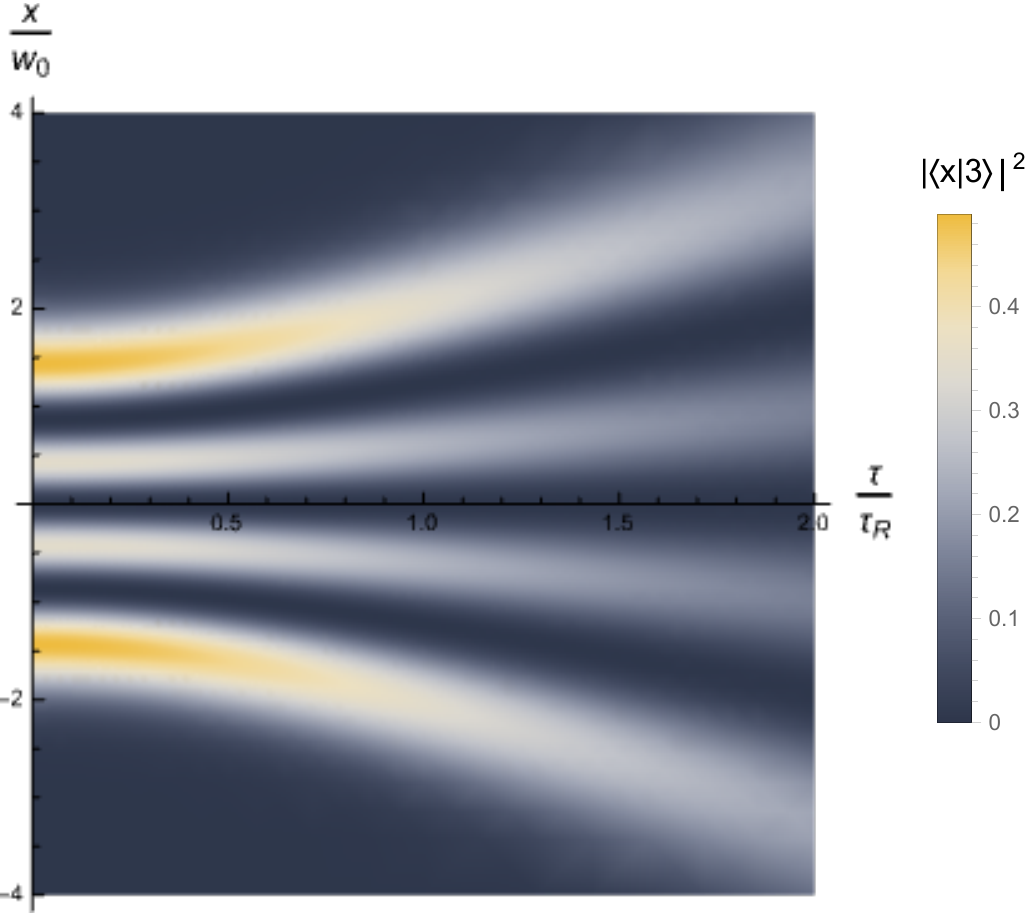}
        &
        \includegraphics[width=.30\linewidth]{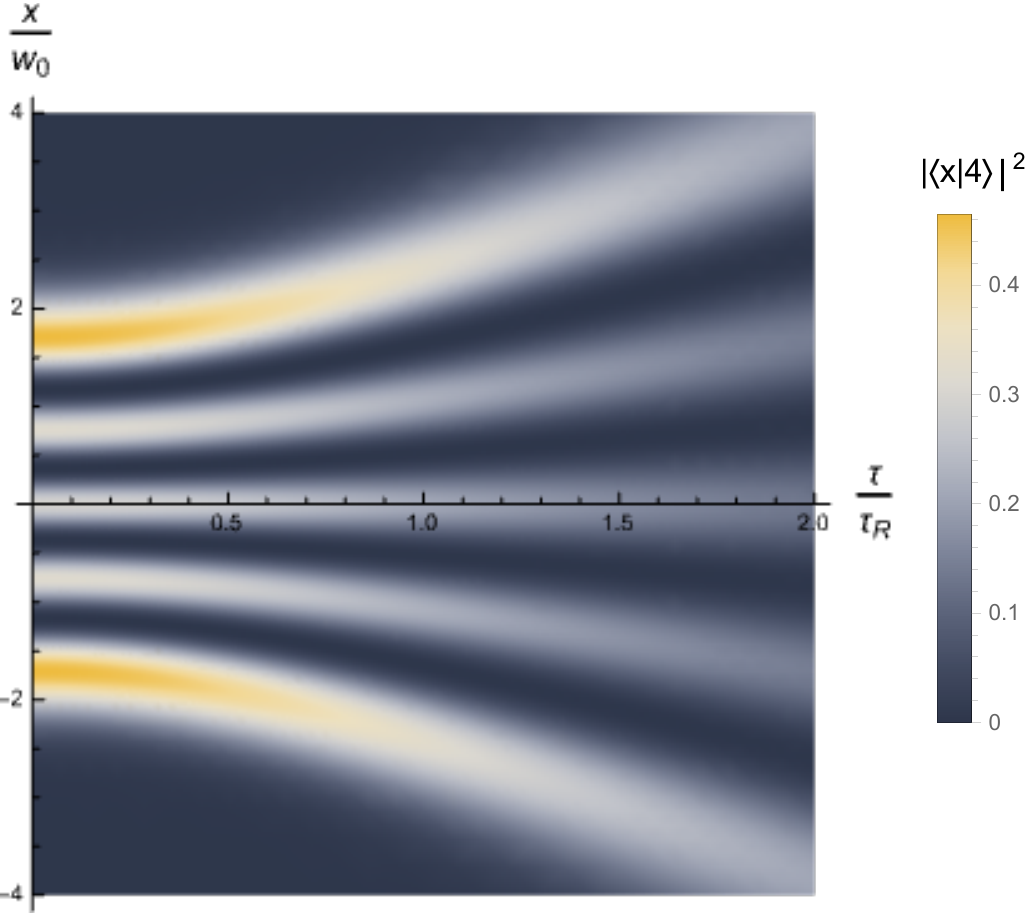}
        &
        \includegraphics[width=.30\linewidth]{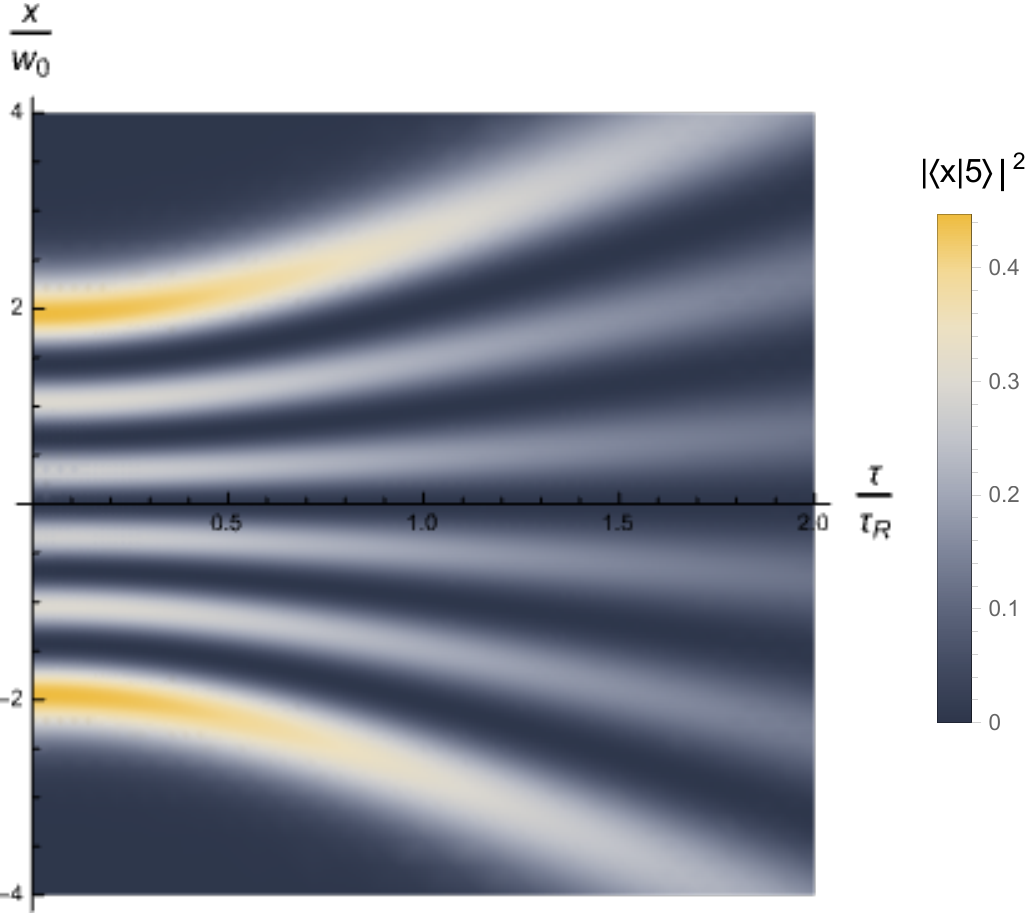}\\
        (e)  & (f)  & (g)  
    \end{tabular}
	\caption{Plot density of the density of probability for the first six modes: (a) n=0, (b) n=1, (c) n=2, (e) n=3, (f) n=4, and (g) n=5. The color shows $|\braket{x}{n}|^2$. The horizontal axis represents the proper time divided by the Rayleigh time given in Eq.~\eqref{RayleighTime}. The perpendicular axis shows the position divided by the initial width. }
	\label{fig:StandardProbabilityDensityPlot}
\end{figure*}

\subsection{Standard Hermite basis}
\label{Standard_basis}
The elegant basis is not orthogonal.  The elegant Hermite modes provide a non-orthogonal complete basis for the physical Hilbert space.  The non-orthogonality can be checked by looking at the inner products of the first few modes:
\begin{equation}
	\begin{array}{c|ccccccccc}
		& \ket{\bar{0}} &  \ket{\bar{1}} & \ket{\bar{2}} &\ket{\bar{3}} &\ket{\bar{4}} & \ket{\bar{5}} &
		\cdots\\
		\hline
		\bra{\bar{0}}	&1 & 0 & -\frac{1}{\sqrt{3}} & 0 & \sqrt{\frac{3}{35}} & 0 & 
		\cdots\\
		\bra{\bar{1}}	&0 & 1 & 0 & -\sqrt{\frac{3}{5}} & 0 & \sqrt{\frac{5}{21}} &
		\cdots\\
		\bra{\bar{2}}	&-\frac{1}{\sqrt{3}} & 0 & 1 & 0 & -\sqrt{\frac{5}{7}} & 0 &
		\cdots \\
		\bra{\bar{3}}	&0 & -\sqrt{\frac{3}{5}} & 0 & 1 & 0 & -\frac{\sqrt{7}}{3} &
		\cdots\\
		\bra{\bar{4}}	&\sqrt{\frac{3}{35}} & 0 & -\sqrt{\frac{5}{7}} & 0 & 1 & 0 &
		\cdots \\
		\bra{\bar{5}}	&0 & \sqrt{\frac{5}{21}} & 0 & -\frac{\sqrt{7}}{3} & 0 & 1  &
		\cdots\\
		\vdots & \vdots & \vdots& \vdots& \vdots& \vdots& \vdots& \ddots
	\end{array}
\end{equation}
In the above table, the element in row $\bra{\bar{i}}$ and column $\ket{\bar{j}}$ represents $\braket{\bar{i}}{\bar{j}}$.
The phase and amplitude are not easily readable in the elegant notation too. It is easier to work with a standard orthogonal basis wherein the phase and amplitude of field are easily readable \cite{pampaloni2004gaussian}. The standard modes can be obtained by constructing a normal basis that we shall represent by  $\ket{n}$ from the elegant basis:
\begin{eqnarray}
	\ket{0} &=& \ket{\bar{0}},\\
	\ket{1} & = & \ket{\bar{1}},\\
	\ket{2} &=& \frac{1}{\sqrt{2}}\ket{\bar{0}}+\sqrt{\frac{3}{2}}\ket{\bar{2}},\\
	\ket{3} & = & \sqrt{\frac{2}{5}} \ket{\bar{3}}-\sqrt{\frac{3}{5}} \ket{\bar{1}},\\
	\ket{4}&=&\frac{1}{2} \sqrt{\frac{3}{2}} \ket{\bar{0}}+\frac{3}{\sqrt{2}}\ket{\bar{2}}+\frac{1}{2} \sqrt{\frac{35}{2}} \ket{\bar{4}},\\
	\ket{5} &= & \sqrt{\frac{5}{21}} \ket{\bar{1}}-\frac{2}{3} \sqrt{\frac{10}{7}} \ket{\bar{3}}+\frac{2}{3} \sqrt{\frac{2}{7}} \ket{\bar{5}}, \\
	\vdots \nonumber \
\end{eqnarray}
The standard modes can also be directly represented in terms of the Hermite functions.  The standard basis defines:
\begin{eqnarray}
	\label{RayleighTime}
	\tau_{\!_R} &=& \frac{1}{2}m w_0^2,\\
\label{widthtau}
	w&=& w_0 \sqrt{1+ \left(\frac{\tau}{\tau_{\!_R}}\right)^2}.
\end{eqnarray}
Note that we are using a compact notation: $w$ is real, and is a function of $\tau$, and $\tau_{\!_R}$ is the Rayleigh time. The width doubles at $\tau=\tau_R$: $w(\tau_{\!_R})=2 w_0$.  The standard basis  defines $\psi$ as the Gouy phase \cite{Gouy}:
\begin{equation}
	\label{Gouy}
	\psi_{n} = -\left(n+\frac{1}{2}\right) \arctan{\left(\frac{\tau}{\tau_{\!_R}}\right)} \,,
\end{equation}
which is a function $\tau$. The \textcolor{black}{$n$th} Hermite-Gaussian mode in the standard basis with an initial width of $w_0$ is then given by:
\begin{eqnarray}
	\label{standardmoden}
	\braket{x}{n}_{w_0}&=&\left(\frac{1}{2\pi}\right)^{\frac{1}{4}}\frac{2^{\frac{-n+1}{2}}}{ \sqrt{n!w}}H_{n}\left(\frac{\sqrt{2}x}{w}\right)\nonumber\\
	&\times&\exp\left(-\frac{ x^2 }{w^2}+\frac{i x^2 \tau}{ w^2\tau_{\!_R}}+ i \psi_{n}\right).
\end{eqnarray}

In the remaining of this section, we drop the subscript of $w_0$ for the sake of simplicity, but we shall recover it in the next section. The standard  Hermite modes provide a complete orthogonal basis for the physical Hilbert space:
\begin{eqnarray}
	\label{completenessS}
	\sum_{n=0}^{\infty} \ket{n}\!\bra{n} &=&1,\\
	\label{orthogonalityS}
	\braket{m}{n} &=& \delta^{mn},
\end{eqnarray}
where $\delta^{mn}$ represents the Kronecker delta, and $\braket{m}{n}=\int_{-\infty}^{\infty} dx \braket{m}{x} \braket{x}{n}$, and $\braket{m}{x}$ is the complex conjugate of $\braket{x}{m}$.  
 
\begin{figure}[t]
 	\centering
 	\includegraphics[width=.90\linewidth]{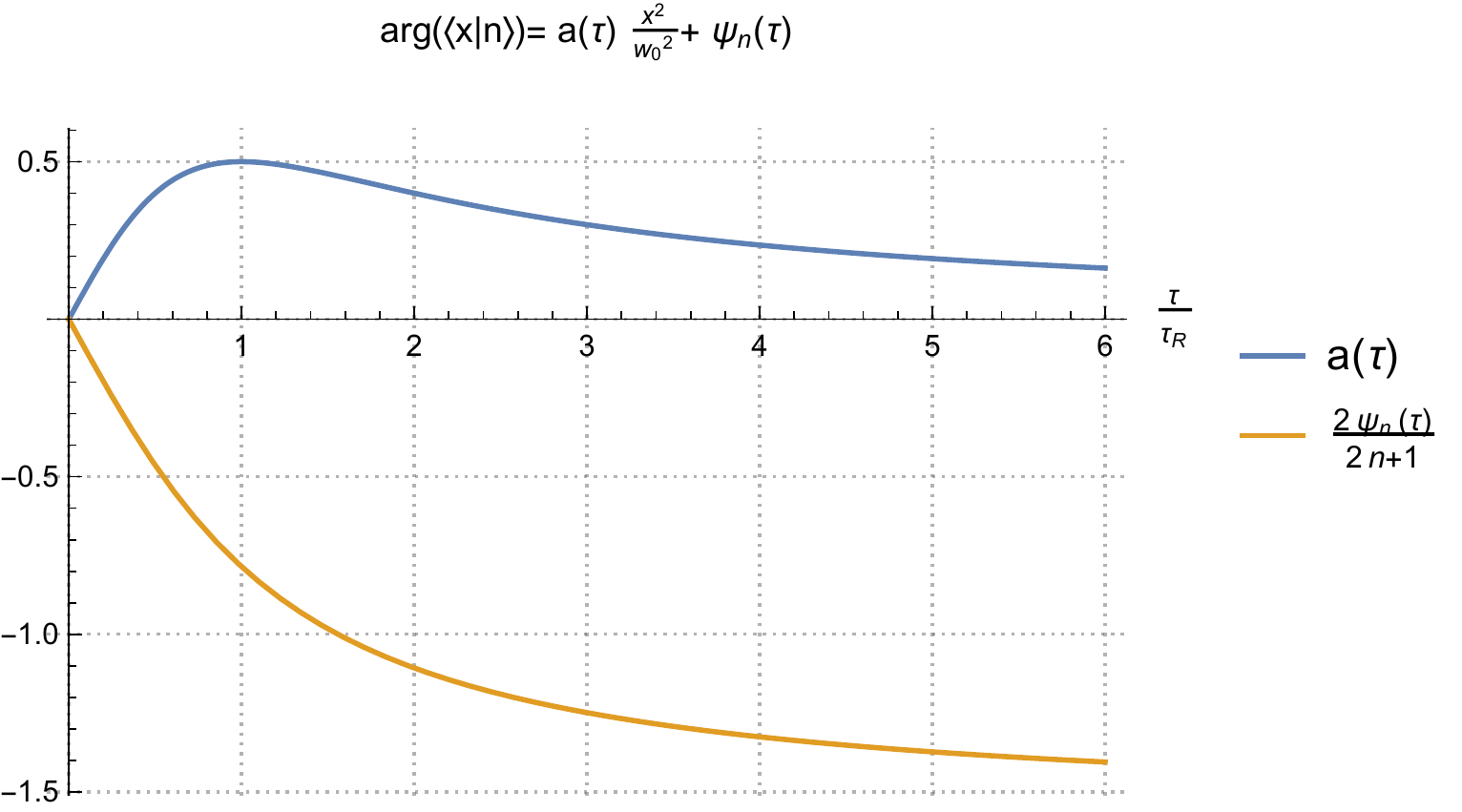}
 	\caption{Phase of the standard Hermite mode n, $\braket{x}{n}$, can be expressed by $\arg\braket{x}{n}= a(\tau) \left(\frac{x^2}{w_0^2}\right)+\psi_n(\tau)$ where $w_0$ is the initial width, $\psi_n$ is the Gouy phase for mode $n$ defined in Eq.~\eqref{Gouy}, and $a(\tau)$ is defined in Eq.~\eqref{a_tau_definition}. Note that $\tau_{\!_R}$ is the Rayleigh time defined in \eqref{RayleighTime}.}
 	\label{Fig:PhasePlot}
 \end{figure}

The density plot of the first six modes is provided in Fig.  \ref{fig:StandardProbabilityDensityPlot}. We observe that $\braket{x}{n}$ in the spacetime continuum is always composed of $n+1$ strips.  The phase of mode $n$ can be re-expressed by:
 \begin{eqnarray}
 	\arg(\braket{x}{n})&=& a(\tau)\, \left(\frac{x^2}{w_0}\right) + \psi_n(\tau),\\
 	\label{a_tau_definition}
 	a(\tau) &=&  \frac{\tau/\tau_{\!_R}}{(1+\tau/\tau_{\!_R})^2}.
\end{eqnarray}
Figure~\ref{Fig:PhasePlot} depicts $a(\tau)$ and $\psi_n(\tau)$ versus $\tau$.  

The completeness of the standard Hermite basis implies that the wave function of the particle in its rest frame, as given in \eqref{PhiField} and \eqref{FieldPhiTilde}, can be written by:
\begin{equation}
	\label{neutrino_wave_restframe}
	\phi(x,\tau) = e^{-i m\tau } \sum_{n=0}^{\infty} d_n \braket{x}{n},
\end{equation}
where $d_n$ are some constant complex numbers. So, let us first look at some physical properties of each mode. The expectation value of the kinetic energy of mode $n$ is given by:
\begin{eqnarray}
	E_n &=& -\frac{1}{2m} \sandwich{n}{\partial_x^2}{n} = \frac{1}{m w_0^2}\left(n+\frac{1}{2}\right).
\end{eqnarray} 
The energy eigenvalues match that of a harmonic oscillator with frequency $\omega=\frac{1}{m w_0^2}$. It also holds:
\begin{eqnarray}
\label{x2naverage}
    \sandwich{n}{x^2}{n} &=& \frac{2n+1}{4} w^2, \\
    \sandwich{n}{p^2}{n} &=& \frac{2n+1}{w_0^2},
\end{eqnarray}
where $w$ is the width of the wave function at the proper time $\tau$ defined in \eqref{widthtau}. Also, notice that it holds: 
\begin{eqnarray}
\label{mntau}
  \sandwich{n}{x^2}{m}= \left.\frac{w^2}{w_0^2}\sandwich{n}{x^2}{m}\right|_{\tau=0}.
\end{eqnarray}
The expectation values of the squared of position and momentum, therefore, satisfy:
\begin{eqnarray}
    \sandwich{n}{p^2}{n}\sandwich{n}{x^2}{n}&=&\frac{(2n+1)^2}{4} \left(1+ \frac{\tau^2}{\tau_{\!_R}^2}\right),
\end{eqnarray}
where $\tau_{\!_R}$ is the Rayleigh time given in Eq.~\eqref{RayleighTime}.
Recalling that $\sandwich{n}{p}{n}=\sandwich{n}{x}{n}\textcolor{black}{=0}$, we see that the zero mode saturates the uncertainty principle at $\tau=0$. Also, notice that for any $m, n$ and any arbitrary integer number of ${\textcolor{black}{\tilde{\alpha}}}$: 
\begin{eqnarray}
\label{propertyHermit}
     \sandwich{n}{x^{\textcolor{black}{\tilde{\alpha}}}}{m} &=& \left(\frac{w}{w_0}\right)^{\textcolor{black}{\tilde{\alpha}}} e^{i(\psi_n-\psi_m)} \left.\sandwich{n}{x^{\textcolor{black}{\tilde{\alpha}}}}{m}\right|_{\tau=0},\nonumber\\
     \sandwich{n}{p^{\textcolor{black}{\tilde{\alpha}}}}{m} &=& e^{i(\psi_n-\psi_m)} \left.\sandwich{n}{p^{\textcolor{black}{\tilde{\alpha}}}}{m}\right|_{\tau=0},
\end{eqnarray}
where $w$ is the width at time $\tau$ defined in Eq.~\eqref{widthtau}, $w_0$ is the initial width, and $\psi_n$ and $\psi_m$ are the Gouy phase for mode $n$ and $m$ defined in Eq.~\eqref{Gouy}. To calculate the right-hand side of Eq.~\eqref{propertyHermit}, one can define a new variable of $y$ by:
\begin{eqnarray}
\label{ytilde_ref}
\textcolor{black}{\tilde{y}} &=& \sqrt{2}\,\frac{x}{w_0},\\
    \braket{\textcolor{black}{\tilde{y}}}{n} &=& \left.\left((\frac{\partial x}{\partial y})^{\frac{1}{2}}\braket{x}{n}\right)\right|_{\tau=0},
\end{eqnarray}
then $\braket{\textcolor{black}{\tilde{y}}}{n}$ satisfies:
\begin{eqnarray}
 \int d\textcolor{black}{\tilde{y}} \abs{\braket{\textcolor{black}{\tilde{y}}}{n}}^2 &=&1,\\
    \left(-\frac{1}{2} \partial_{\textcolor{black}{\tilde{y}}}^2 + \frac{1}{2} {\textcolor{black}{\tilde{y}}}^2\right)\braket{\textcolor{black}{\tilde{y}}}{n} &=& \left(n+\frac{1}{2}\right)\braket{\textcolor{black}{\tilde{y}}}{n},
\end{eqnarray}
which can be perceived as the  Hamiltonian of a simple harmonic oscillator with $m=\omega=1$. We can define the ladder operators of \cite{PhysRevA.48.656}: 
\begin{eqnarray}
    a^\dagger &=& \frac{-\partial_{\textcolor{black}{\tilde{y}}}+\textcolor{black}{\tilde{y}}}{\sqrt{2}},\\
    \label{a_anihilation_referee}
    a &=& \frac{\partial_{\textcolor{black}{\tilde{y}}}+{\textcolor{black}{\tilde{y}}}}{\sqrt{2}},
\end{eqnarray}
that satisfy $[a,a^\dagger]=1$ to alter the mode number:
\begin{eqnarray}
    a^\dagger \ket{n} &=& \sqrt{n+1} \ket{n+1},\\
    a \ket{n+1} &=& \sqrt{n} \ket{n}.
\end{eqnarray}
which are useful to obtain the sandwich of $x$ or $p$ between two modes on the right-hand side of Eq.~\eqref{propertyHermit} by utilising:
\begin{eqnarray}
    x &=& \frac{w_0}{2}(a+a^\dagger),\\
    p &=& \frac{i(a^\dagger-a)}{w_0}.
\end{eqnarray}

\textcolor{black}{In the following we would like} to estimate the systematic error in ignoring second derivative of $\tau$ in \eqref{ParaxialApproximation}, let us multiply both sides of  \eqref{ParaxialApproximation} with $\tilde{\phi}^*_n$ and integrate it over $x$ from $x=-\infty, \infty$:
\begin{equation}
	\label{nUpperLimit}
	\sandwich{n}{(i\partial_\tau)^2}{n} + 2m \sandwich{n}{i\partial_\tau}{n}\approx  2m \sandwich{n}{i\partial_\tau}{n} .
\end{equation}
The systematic error in this approximation is  given by 
\begin{equation}
	\text{Error}(n) = \frac{\sandwich{n}{H^2}{n}}{\sandwich{n}{H^2}{n}+ 2m \sandwich{n}{H}{n}},
\end{equation}
where $i\partial_\tau = H$ is utilized \textcolor{black}{and}  $H= -\frac{1}{2m}\partial_x^2$ represents the Hamiltonian. Using the explicit form of $\braket{x}{n}$ yields:
\begin{eqnarray}
	 \sandwich{n}{H}{n} &=& E_n,\\
	\sandwich{n}{H^2}{n} &=& \frac{6n^2+6n +3}{4m^2 w_0^4},
\end{eqnarray}
which can be utilised to write: 
\begin{eqnarray}
\label{ErrorAux}
	\frac{1}{\text{Error}(n)}= 1+ \frac{4 m^2 w_0^2}{3}  \left(1- \frac{n^2}{n^2+n+\frac{1}{2}}\right).
\end{eqnarray} 
Any experiment or measurement is performed with a certain precision. The systematic error should be always smaller than the precision, represented by $p_r$:
\begin{equation}
	\text{Error}(n) < p_r, 
\end{equation} 
The precision is a very small positive real number: $p_r \ll 1$. So \eqref{ErrorAux} can be rewritten by:
\begin{eqnarray}
	\frac{n^2}{n^2 + n + \frac{1}{2}}< 1- \frac{3}{4m^2 w_0^2}\left(\frac{1}{p_r}\right).
\end{eqnarray} 
For any given precision and initial width, the right-hand side of the above inequality is smaller than $1$, but the limit of $n\to \infty$ of the right-hand side is $1$. This proves that for any given width and precision, there exists an upper bound on $n$ where the systematic error becomes larger than the precision of the experiment. Assuming that the upper bound is large, it can be shown that:
\begin{eqnarray}
    n &< & n_{{\!}_{\text{ub}}},\\
  n_{{\!}_{\text{ub}}} &=& \frac{4}{3} p_r \, m^2 w_0^2,
\end{eqnarray}
The upper bound of $n_{{\!}_{\text{ub}}}$ is where the relativistic corrections in the rest frame of the particle become important. We assume that $m w_0$ is very large, and the precision is such that $1\ll n_{{\!}_{\text{ub}}}$, and we consider only modes of $n\ll n_{{\!}_{\text{ub}}}$.  We, therefore, consistently ignore the relativistic corrections to the Schr\"odinger equation in the rest frame of the wave function. 

\subsection{Localised solutions in (3+1)-dimensions}
\label{Section:HermitModesin3D}
Equation~\eqref{standardmoden} can be utilized to write $\langle x_1|n_1\rangle_{w_{01}}$ , $\langle x_2|n_2\rangle_{w_{02}}$ and $\langle x_3|n_3\rangle_{w_{03}}$ respectively as the $n_1$, $n_2$ and $n_3$ Hermite mode for $(x^1, \tau)$, $(x^2, \tau)$ and $(x^3, \tau)$ coordinates  with the initial width of $w_{01}$ , $w_{02}$ and $w_{03}$. They satisfy:
\begin{eqnarray}
    - i \partial_\tau \langle x_i|n_i\rangle_{w_{0i}} &=& \frac{1}{2m} \partial_{x_i}^2 \langle x_i|n_i\rangle_{w_{0i}},
\end{eqnarray}
where $i\in\{1,2,3\}$.  Let us define:
\begin{eqnarray}
\label{3DHermitMode}
    \langle \vec{x}|\vec{n}\rangle_{\vec{w}_0} = \prod_{i=1}^3\langle x_i|n_i\rangle_{w_{0i}}, 
\end{eqnarray}
where $\vec{x}=(x_1,x_2,x_3)$, $\vec{n}=(n_1,n_2,n_3)$ and $\vec{w}_0=(w_{01},w_{02},w_{03})$, respectively. 
Then, $\langle \vec{x}|\vec{n}\rangle_{\vec{w}_0}$ satisfies the Schr\"odinger equation in $(3+1)$-dimensions:  
\begin{equation}
     i \partial_\tau  \langle \vec{x}|\vec{n}\rangle_{\vec{w}_0} = -\frac{\nabla^2}{2m} \langle \vec{x}|\vec{n}\rangle_{\vec{w}_0},
\end{equation}
where: 
\begin{eqnarray}
\label{Nabla2}
     \nabla^2=\sum_{i=1}^3\left(\frac{\partial}{\partial x_i}\right)^2.
\end{eqnarray}
Notice that $\vec{w}_0$ represent the initial width for $x_1$, $x_2$ and $x_3$ coordinates. For any choices of $\vec{w}_0$, $\langle \vec{x}|\vec{n}\rangle_{\vec{w}_0}$ provides a complete orthogonal basis for \textcolor{black}{localized solutions of} the Schr\"odinger equation in $(3+1)$-dimensions. Here, we choose the same initial width of $w_0$ for all the coordinates and set:
\begin{equation}
\label{w0to3}
    w_0=w_{01}=w_{02}=w_{03}.
\end{equation} 
For the sake of simplicity, we also drop subscript $\vec{w}_0$. So the completeness of the basis can be succinctly expressed in Dirac notation:
\begin{eqnarray}
\label{StandardHermitBasis}
    \sum_{\vec{n}} |\vec{n}\rangle \langle \vec{n}| &=& 1,\\
    \langle \vec{n}|\vec{m}\rangle &=& \delta_{\vec{n},\vec{m}},
\end{eqnarray}
where: 
\begin{eqnarray}
\label{delta3D}
    \delta_{\vec{n},\vec{m}}= \prod_{i=1}^3 \delta_{n_i,m_i}.    
\end{eqnarray}
We define the zero state or the ground state by $\vec{0}=(0,0,0)$, and represent it by $\ket{0}=|{\vec{0}}\rangle$. We call the state $\ket{\vec{n}}$ as the excitation with number $n_1$, $n_2$ and $n_3$ in respectively the directions of $x^1$, $x^2$ and $x^3$

Any \textcolor{black}{localized} solution to the Schr\"odinger equation can be represented by:
\begin{equation}
    |\tilde{\phi}\rangle= \sum_{\vec{n}} c_{\vec{n}} |\vec{n}\rangle,
\end{equation}
where $|\vec{n}\rangle$ represents the localised Hermite mode of $\vec{n}$, and $c_{\vec{n}}$ are complex constant numbers identified by the initial conditions. Recall that we are working in the rest frame of the wave function, so the expectation of the linear momentum must vanish:
\begin{eqnarray}
\label{vanishingP}
  \sandwich{\tilde{\phi}}{\vec{p}\,}{\tilde{\phi}} = 0.
\end{eqnarray}
We further can choose the origin of the coordinates by demanding:
\begin{eqnarray}
\label{vanishingX}
  \sandwich{\tilde{\phi}}{\vec{x}\,}{\tilde{\phi}} = 0.
\end{eqnarray}
Let us call $\vec{n}$ and $\vec{m}$ as two immediate neighbours if $\abs{n_1-m_1}= 1$ or $\abs{n_2-m_2}= 1$ or $\abs{n_3-m_3}= 1$.  Utilizing $2 x H_n(x) = H_{n+1}(x)+ 2 n H_{n-1}(x)$ or by direct calculation, it can be proven that \eqref{vanishingP} and \eqref{vanishingX} hold true when $c_{\vec{n}}$ vanishes for at least one of any immediate neighbours. 

\section{Generally curved spacetime geometry}
\label{Section:curvedgeometry}
Fermi coordinates can be utilised to represent the spacetime geometry in the vicinity of a classical \textcolor{black}{observable} in the rest frame of its localised wave function. \textcolor{black}{We consider that} \textcolor{black}{the observable} moves on a time-like geodesic.  
The expansion of Fermi coordinates adapted to the time-like geodesic $\gamma$, up to the quadratic transverse directions, are given by Manasse and Misner \cite{Manasse:1963zz}. The expansion up to the fourth order is given in \cite{doi:10.1063/1.524292}. To reproduce this expansion, we first introduce $\epsilon$ as a small dummy parameter to systematically track the perturbation. At the end of the perturbative calculation, we set $\epsilon=1$.  In the expansion given in \cite{doi:10.1063/1.524292}, we scale the transverse direction by the factor of $\epsilon$.  We present the expansion to the line element up to the third order by:
\begin{eqnarray}
	\label{Eq4.1}
	ds^2 &=& {}^{(0)}ds^2 + {}^{(2)}ds^2 \epsilon^2 +  {}^{(3)}ds^2 \epsilon^3 + O(\epsilon^4),    
\end{eqnarray}
Notice that we have used the left upper script of ${}^{(n)}$ to represent the $n^\text{th}$ order correction to the line element.  And notice that by definition, there is no perturbation at the linear order of $\epsilon$, because by definition, the first derivative of the metric in the Fermi coordinates evaluated along the central geodesic vanishes. The explicit corrections up the third order are given by: 
\begin{eqnarray}
	{}^{(0)}ds^2 &=& -(dx^0)^2 + \sum_{i=1}^3  (dx^i)^2, \\
	{}^{(2)}ds^2&=& - R_{0l0m} x^l x^m (dx^0)^2 \nonumber\\&-& \frac{4}{3} R_{0 l i m} x^l x^m dx^0 dx^i\nonumber\\&-&  \frac{1}{3} R_{iljm} x^l x^m  dx^i dx^j\,,\\
	{}^{(3)}ds^2 &=& -\frac{1}{3} R_{0l0m;n} x^l x^m x^n (dx^0)^2
	\nonumber \\
	&-& \frac{1}{2} R_{0 l i m;n} x^l x^m x^n dx^0 dx^i\nonumber\\
	&-& \frac{1}{6} R_{i l j m;n} x^l x^m x^n dx^i dx^j,
\end{eqnarray}
where $R_{\mu\alpha\beta\nu}$ represents the components of the Riemann tensor computed along the time-like geodesics, $R_{\mu\alpha\beta\nu;\gamma}=\nabla_\gamma$ represents components of \textcolor{black}{the} covariant derivative of the Riemann tensor computed along the time-like geodesics, $x^0$ is the proper time of the \textcolor{black}{observable}, and $x^i$ are the spatial transverse directions to the time-like geodesic. Notice that Manasse and Misner's work~\cite{Manasse:1963zz} with a non-standard sign convention for the Riemann tensor: ${}^\text{(Manasse-Misner)} R_{\nu~\rho\sigma}^{~\mu}= {}^\text{(MTW)} R_{~\nu\rho\sigma}^{\mu}$, where ${}^\text{(MTW)} R_{~\nu\rho\sigma}^{\mu}$ is the standard sign convention in GR textbook by Misner, Thorne and Wheeler \cite{Misner:1973prb} which is the most widely used sign convention nowadays. The standard sign convention is utilised in Eq.~\eqref{Eq4.1}. Also notice that $dx^{0} dx^i$ in line $4$ of Eq. (45) in \cite{doi:10.1063/1.524292} is missing an overall factor of $2$ that is included in Eq.~\eqref{Eq4.1}.

The line element evaluated at $x^a=0$ coincides with that of Minkowski spacetime geometry:
\begin{eqnarray}
    ds^2|_{x^a=0} = {}^{(0)}ds^2.
\end{eqnarray}
We call the time-like geodesic defined by $x^a=0$ as the central time-like geodesic. 

\subsection{Residual Newtonian gravity in the quadratic order corrections}
To define quantum mechanics in \textcolor{black}{a generally} curved spacetime geometry, we follow the approach of ref. \cite{Exirifard:2021cav}, and in this section, we adopt its outcome that the leading correction to the Schr\"odinger equation in the flat spacetime geometry is given by: 
\begin{eqnarray}
	\label{SchPer}
	i  \partial_0 \Psi = \left(-\frac{\nabla^2}{2m}+\epsilon^2 \frac{m}{2} R_{0a0b} x^a x^b \right) \Psi+O(\epsilon^3).
\end{eqnarray}
where $\nabla^2$ is defined in \eqref{Nabla2}, while the vacuum expectation of any \textcolor{black}{observable} $\hat{O}$ coincides with that of the flat spacetime geometry: 
\begin{equation}	
	\label{Eq5.21}
	\langle \hat{O} \rangle =  \int d^3x\,  \Psi^{*} \hat{O} \Psi,
\end{equation}
\textcolor{black}{Notice that we are considering low energy wave functions that can be expanded in terms of finite number of the Gaussian Hermite modes given in \eqref{standardmoden}, and as these modes are exponentially localized around $x^a=0$ with width much smaller than the curvature of the spacetime geometry, the boundary of integration in  \eqref{Eq5.21} can be consistently extended to infinity. }
Note that $(\frac{1}{2} R_{0a0b} x^a x^b)$ which appears in Eq. \eqref{SchPer} is the residual Newtonian potential and is equal to $\frac{1+g_{00}}{2}$ -- here, $g_{00}$ is the zero-zero component of the metric presented in Eq. \eqref{Eq4.1}.  Reference~\cite{Exirifard:2021cav} has proven that the wave function can be scaled such that Eq.~\eqref{Eq5.21} holds, and the contribution of the rest of the components of the metric can be neglected at the leading order. We write a perturbative expansion for the wave function:
\begin{equation}
\label{PsiEpsilon}
    \Psi = \Psi^{(0)} + \epsilon^2 \Psi^{(2)} + O(\epsilon^3),
\end{equation}
which can be substituted into Eq.~\eqref{SchPer}:
\begin{eqnarray}
\label{Psi0Eq}
    \left(i  \partial_0  +\frac{\nabla^2}{2m}\right) \Psi^{(0)} &=& 0,\\
    \label{Psi1Eq}
    \left(i  \partial_0  +\frac{\nabla^2}{2m}\right) \Psi^{(2)} &=& \frac{m}{2} R_{0a0b} x^a x^b \Psi^{(0)},
\end{eqnarray}
where $R_{0a0b}$ is a function of $x^0=\tau$, and $\tau$ is the proper time. We would like to solve Eq. \eqref{Psi1Eq} for any $\Psi^{(0)}$ that satisfies Eq. \eqref{Psi0Eq}. We choose the boundary condition of:
\begin{eqnarray}
\label{BoundaryCondition}
    \left.\Psi^{(2)}\right|_{\tau=0} = 0.
\end{eqnarray}
In analogy with \cite{Exirifard:2020yuu}, let us first define:
\begin{eqnarray}
\label{R2functions}
    {\,}^{1}\!{\cal R}_{ab}(\tau) &=& \int_0^{\tau} d\tilde{\tau} R_{0a0b}(\tilde{\tau}),\nonumber\\
    {\,}^{2}\!{\cal R}_{ab}(\tau) &=& \int_0^{\tau} d\tilde{\tau}{{\,}^{\textcolor{black}{1}}}\!{\cal R}_{ab}(\tilde{\tau}),\nonumber\\
    {\,}^{3}\!{\cal R}_{ab}(\tau) &=& \int_0^{\tau} d\tilde{\tau}{{\,}^{2}\!{\cal R}}_{ab}(\tilde{\tau}),
\end{eqnarray}
which are symmetric in their indices, and they vanish at $\tau=0$. We refer to Eq.~\eqref{R2functions} as ${\cal R}_2$ functions.  The Einstein equation of motion for the background spacetime geometry in the vacuum implies that the Ricci tensor vanishes. Some components of the Ricci tensor evaluated on the central time-like geodesic are given by:
\begin{eqnarray}
\label{R00}
    0=R_{00}&=& \delta^{ab} R_{0a0b},\\
\label{R0c}
    0=R_{0c}&=& \delta^{ab} R_{0acb},
\end{eqnarray}
which implies ${\,}^{n}{\cal R}_{ab} \delta^{ab}=0$ \textcolor{black}{for $n\in \{1,2,3\}$} . Let us expand $\Psi^{(2)}$:
\begin{eqnarray}
\label{ansatz1}
    \Psi^{(2)} = {\,}^{1}{\Psi}^{(2)} + \frac{m}{2 i}{\,}^{1}\!{\cal R}_{ab} x^a x^b \Psi^{(0)}, 
\end{eqnarray}
where ${\,}^{1}{\Psi}^{(2)}$ is a general function of $\tau$ and $x^a$. Substituting Eq.~\eqref{ansatz1} into Eq.~\eqref{Psi1Eq} yields:
\begin{equation}
\label{EqPsiTilde}
    \left(i  \partial_0  +\frac{\nabla^2}{2m}\right) {\,}^{1}{\Psi}^{(2)} =   i {\,}^{1}\!{\cal R}_{ab} x^a \partial^b \Psi^{(0)}.
\end{equation}
Now let ${\,}^{1}{\Psi}^{(2)}$ be expressed by: 
\begin{eqnarray}
\label{anstaz2}
    {\,}^{1}{\Psi}^{(2)} = {{\,}^{2}{\Psi}}^{(2)} +   {{\,}^{2}\!{\cal R}}_{ab} x^a \partial^b \Psi^{(0)},
\end{eqnarray}
where ${{\,}^2\Psi}^{(2)}$ is a general function of $\tau$ and $x^a$. Employing Eq.~\eqref{anstaz2} in Eq.~\eqref{EqPsiTilde} yields:
\begin{eqnarray}
    \left(i  \partial_0  +\frac{\nabla^2}{2m}\right)  {{\,}^{2}{\Psi}}^{(2)} = -\frac{1}{m} {{\,}^{2}\!{\cal R}}_{ab} \partial^a \partial^b \Psi^{(0)},
\end{eqnarray}
which is solved by:
\begin{eqnarray}
\label{anstaz3}
    {{\,}^{2}{\Psi}}^{(2)} =  \frac{ i}{m} {\,}^{3}\!{\cal R}_{ab} \partial^a \partial^b \Psi^{(0)}.
\end{eqnarray}
Thus, from Eqs \eqref{ansatz1}, \eqref{anstaz2} and \eqref{anstaz3}, $ \Psi^{(2)}$ is given by:
\begin{equation}
\label{Psi2Sol}
    \Psi^{(2)}=\left(\frac{m}{2 i}{\,}^{1}\!{\cal R}_{ab} x^a x^b+  {{\,}^{2}\!{\cal R}}_{ab} x^a \partial^b+ \frac{i}{m} {\,}^{3}\!{\cal R}_{ab} \partial^a \partial^b\right)\Psi^{(0)},  
\end{equation}
which is a solution to Eq. \eqref{Psi1Eq} for any $\Psi^{(0)}$ that solves \eqref{Psi0Eq}, and satisfies the boundary condition chosen in \eqref{BoundaryCondition}. It is more convenient to define the following Hermitian operators: 
\begin{eqnarray}
\label{QOperator}
Q^{(2)} &=&  -\frac{m}{2}{\,}^{1}\!{\cal R}_{ab} x^a x^b+ {{\,}^{2}\!{\cal R}}_{ab} x^a p^b -\frac{1}{m} {\,}^{3}\!{\cal R}_{ab} p^a p^b,\nonumber\\
\end{eqnarray}
where $p^a=-i\partial^a$ represents the momentum.  In the Dirac notation, one can write:
\begin{eqnarray}
    \label{PsiQ}
     \ket{\Psi}  &=& \exp\left(i Q\right) |\Psi^{(0)}\rangle ,\\
     \label{QepsilonExpansion}
     Q&=&  \epsilon^2 Q^{(2)} +  \epsilon^3 Q^{(3)}+ \cdots.
\end{eqnarray}
It can be easily seen that the norm is left intact by the $\epsilon$ perturbation:
\begin{eqnarray}
     \braket{\Psi}{\Psi}  =\langle{\Psi^{(0)}}|{\Psi^{(0)}}\rangle + O(\epsilon^3),
\end{eqnarray}
which is due to the fact that $Q^{(2)}$ is Hermitian: $Q^{(2)}=Q^{(2)\dagger}$.

\subsubsection{Cross-talk coefficients}
Consider the case that $|{\Psi^{(0)}}\rangle$ coincides with a Hermite mode at $\tau=0$: $|{\Psi^{(0)}}\rangle = \ket{\vec{n}}$,
where $\ket{\vec{n}}$ is defined in Eq.~\eqref{3DHermitMode}. The wave function in the proper time $\tau$ is given by 
$\ket{\Psi}  = \exp\left( i \epsilon^2 Q^{(2)}\right) \ket{\vec{n}}$.
The state $\ket{\Psi}$, per se, can be expressed in terms of the Hermite modes: $\ket{\Psi}  = \sum_{\vec{m}} C_{\vec{m},\vec{n}} \ket{\vec{m}}$. We refer to $C_{\vec{m},\vec{n}}$ as the cross-talk coefficients. They encode the change of one mode to another one. At the linear order perturbation, they are given by 
\begin{eqnarray}
     C_{\vec{m},\vec{n}}&=& \delta_{\vec{m},\vec{n}} 
     - \frac{i m}{2}{\,}^{1}\!{\cal R}_{ab}\sandwich{\vec{m}}{x^a x^b}{\vec{n}}\nonumber\\
     &+&i {{\,}^{2}\!{\cal R}}_{ab}\sandwich{\vec{m}}{x^a p^b}{\vec{n}}\nonumber\\
    &-&\frac{i}{m}{\,}^{3}\!{\cal R}_{ab}\sandwich{\vec{m}}{p^a p^b}{\vec{n}},
\end{eqnarray}
where we set $\epsilon=1$. Due to the properties of the standard Hermite modes reviewed in Eq.~\eqref{propertyHermit}, the cross-talk coefficients can be further simplified to: 
\begin{eqnarray}\label{C1}
     C_{\vec{m},\vec{n}}&=& \delta_{\vec{m},\vec{n}} 
     - \frac{i m w^2}{2 w_0^2} {\,}^{1}\!{\cal R}_{ab} e^{i(\psi_n-\psi_m)}
     \left.\sandwich{\vec{m}}{x^a x^b}{\vec{n}}\right|_{\tau=0}\nonumber\\
     &+&\frac{iw}{w_0} e^{i(\psi_n-\psi_m)} {{\,}^{2}\!{\cal R}}_{ab}
     \left.\sandwich{\vec{m}}{x^a p^b}{\vec{n}}\right|_{\tau=0}\nonumber\\
    &-&\frac{i}{m} e^{i(\psi_n-\psi_m)}{\,}^{3}\!{\cal R}_{ab}\left.\sandwich{\vec{m}}{p^a p^b}{\vec{n}}\right|_{\tau=0},
\end{eqnarray}
where Eq. \eqref{delta3D} is employed, the expectation value of all values (``moments'') is calculated at $\tau=0$,  $w$ represents the width at the proper time $\tau$ as written in Eq.~\eqref{widthtau},  $w_0$ is the initial width, and $\psi_n$ and $\psi_{\textcolor{black}{m}}$ are Gouy phases given in Eq.~\eqref{Gouy}, respectively. Note that the cross-talk coefficients are a function of the proper time. The dependency on the time is given by the coefficients on the right-hand side of Eq.~\eqref{C1}. 

\subsubsection{Residual net force}
\label{Section:residualforce}
Newton's {\it Lex Secunda} can be utilised to define  the net force acting on the wave function $\Psi$ as the time rate of the change of its average linear momentum:
\begin{eqnarray}
\label{residaulForce}
    \vec{F}&=& \partial_\tau \sandwich{\Psi}{\vec{p}}{\Psi}.
\end{eqnarray}
We can employ Eq.~\eqref{PsiQ} and write:
\begin{eqnarray}
    \vec{F}= \partial_\tau \langle{\Psi^{(0)}}|{e^{-i Q} \vec{p}e^{i Q} }|{\Psi^{(0)}}\rangle,
\end{eqnarray}
where it also assumed that $Q=Q^\dagger$. Utilising the $\epsilon$ expansion series for $Q$ yields a perturbative series for $\vec{F}$:
\begin{eqnarray}
    \vec{F}=0+ \epsilon^2 \vec{F}^{(2)} + \epsilon^3 \vec{F}^{(3)} + O(\epsilon^4). 
\end{eqnarray}
There exists no force acting on the wave function at the zero and linear orders since a free propagating wave function experiences no force in the flat spacetime geometry, and the $\epsilon$ correction vanishes at the linear order. We can use Eq.~\eqref{PsiQ} to write the component of the linear correction to net force:
\begin{eqnarray}
    F^{(2)}_a&=& i \partial_\tau \langle\psi^{(0)}|[p_a,Q^{(2)}]|\psi^{(0)}\rangle,\\
    \label{F3}
    F^{(3)}_a&=& i \partial_\tau \langle\psi^{(0)}|[p_a,Q^{(3)}]|\psi^{(0)}\rangle,
\end{eqnarray}
Recalling the quantum commutation relations between position and linear momentum: 
\begin{eqnarray}
     [x^a,x^b] &=& 0,\\
    {[p^a, p^b]} &=& 0,\\
    {[x^a, p^b]} &=& i \delta^{ab},
\end{eqnarray}
and utilising \eqref{QOperator}, the net force at the second order is simplified to:
\begin{equation}
     F^{(2)}_a =  \partial_\tau \langle\psi^{(0)}| \left(- m {\,}^{1}\!{\cal R}_{ab} x^b+{{\,}^{2}\!{\cal R}}_{ab}p^b \right)|\psi^{(0)}\rangle,
\end{equation}
where Eqs.~\eqref{vanishingP} and \eqref{vanishingX} can be used to conclude:
\begin{eqnarray}
     \vec{F}^{(2)} = 0. 
\end{eqnarray}
This proves that, at the dominant term in the second order level perturbation, the interaction of the curvature of spacetime geometry and the wave function generates no net force acting on an arbitrary freely falling localised wave function. In this order of perturbation, the minor terms generate no net force in the vacuum since they are quadratic in $x^{a}$.  This demonstrates that the Einstein equivalence principle holds for a quantum mechanical system in a non-uniform gravitational field up to the leading gravitational correction.

When one goes beyond the leading correction, the $Q^{(3)}$ operator in Eq.~\eqref{PsiQ} is not guaranteed to remain as a polynomial of second degree in $x^a$, so the expectation value of its communication with $p^a$ will no longer be guaranteed to be vanished by  Eqs.~\eqref{vanishingP} and \eqref{vanishingX}, and the residual force at the third order is not expected to vanish.

\subsection{Residual Newtonian gravity at the cubic order corrections}
\label{secion:CubicCorrections}
We notice that the zero-zero component of the metric in Eq.~\eqref{Eq4.1} is given by:
\begin{equation}
\label{g00}
     g_{00}= -1  - R_{0l0m} x^l x^m \epsilon^2  -\frac{1}{3} R_{0l0m;n} x^l x^m x^n \epsilon^3 + O(\epsilon^4).
\end{equation}
A particle with mass $m$ that moves with a slow velocity in this geometry experiences an effective Newtonian potential given by:
\begin{eqnarray}
     V_{\text{N}} &=& - \frac{g_{00}+1}{2}.
\end{eqnarray}
The Schr\"odinger equation for the wave function of this particle then follows:
\begin{eqnarray}
     i \partial_\tau \Psi = \left(-\frac{\nabla^2}{2m} + m V_{\text{N}}\right) \Psi,
\end{eqnarray}
and the expectation values are \textcolor{black}{observable} and given by Eq.~\eqref{Eq5.21}. Using Eq.~\eqref{g00}, we have,
\begin{eqnarray}
\label{Sch3}
      i \partial_\tau \Psi &=& -\frac{\nabla^2}{2m} \Psi + \epsilon^2 \frac{m}{2} R_{0a0b} x^a x^b \Psi \nonumber\\
      &+&  \epsilon^3 \frac{m}{6} R_{0a0b;c} x^a x^b x^c \Psi + O(\epsilon^4).
\end{eqnarray}
At the leading order, Eq. \eqref{Sch3} is in agreement with Eq. \eqref{SchPer}, which has been derived in \cite{Exirifard:2021cav} by studying quantum field theory in a general curved spacetime geometry. We directly have checked that taking into account the sub-leading corrections in the quantum field theory for a massive scalar field in a general curved spacetime geometry leads to Eq.~\eqref{Sch3}. 

We would like to solve Eq.~\eqref{Sch3} in a perturbative fashion. Thus, we extend the perturbative series for $\Psi$ presented in Eq.~\eqref{PsiEpsilon} to $\epsilon^3$:
\begin{equation}
\label{PsiEpsilon3}
    \Psi = \Psi^{(0)} + \epsilon^2 \Psi^{(2)}+ \epsilon^3 \Psi^{(3)} + O(\epsilon^4).
\end{equation}
Note that $\Psi^{(0)}$ satisfies Eq. \eqref{Psi0Eq} and $\Psi^{(2)}$ is expressed in Eq. \eqref{Psi2Sol}. Employing Eq. \eqref{PsiEpsilon3}  into Eq. \eqref{Sch3} yields:
\begin{eqnarray}
\label{Psi3Eq}
\left(i  \partial_0  +\frac{\nabla^2}{2m}\right) \Psi^{(3)} &=& \frac{m}{6} R_{0a0b;c} x^a x^b x^c \Psi^{(0)}.
\end{eqnarray}
We would like to solve Eq.~\eqref{Psi3Eq} for any $\Psi^{(0)}$ that satisfies Eq. \eqref{Psi0Eq}. We choose the following boundary condition on  $\Psi^{(3)}$:
\begin{eqnarray}
\label{BoundaryCondition3}
    \left.\Psi^{(3)}\right|_{\tau=0} = 0.
\end{eqnarray}
We would like to find part of the correction to the wave function that may generate a non-zero residual force at the order of $\epsilon^3$.  To this aim, we first define the following quantities: 
\begin{eqnarray}
\label{Gabc}
    {\,}^{1}\!{\cal R}_{abc}(\tau) &=& \int_o^\tau d\tilde{\tau} R_{0(a0b;c)}(\tilde{\tau}),
    \nonumber\\
    {{\,}^{2}\!{\cal R}}_{abc}(\tau) &=& \int_o^\tau d\tilde{\tau} {\,}^{1}\!{\cal R}_{abc}(\tilde{\tau}), \nonumber\\
    {\,}^{3}\!{\cal R}_{abc}(\tau) &=& \int_o^\tau d\tilde{\tau} {{\,}^{2}\!{\cal R}}_{abc}(\tilde{\tau}),\nonumber\\  
    {\,}^{4}\!{\cal R}_{abc}(\tau) &=& \int_o^\tau d\tilde{\tau}{\,}^{3}\!{\cal R}_{abc}(\tilde{\tau}),
\end{eqnarray}
where $R_{0(a0b;c)}=\frac{1}{3}(R_{0a0b;c}+R_{0a0c;b}+R_{0c0b;a})$ is symmetric under any permutation of $a,b$, and $c$. We refer to the above quantities as ${\cal R}_3$ functions. Notice that ${\,}^{n}\!{\cal R}_{abc}$ is a function of the Riemann tensor evaluated and integrated over the central time-like geodesic, is symmetric under permutation of their indices, and vanishes at $\tau=0$. 

Equations \eqref{R00} and \eqref{R0c} imply that on the central geodesic, we have,
\begin{eqnarray}
\label{DR00}
    0 &=& R_{00;c} = \delta^{ab} R_{0a0b;c},\\ \nonumber
    0 &=& R_{0c,0} = \delta^{ab} R_{0acb;0}.
\end{eqnarray}
The second Bianchi identity evaluated on the central time-like geodesic and contracted by $\delta^{ab}$ implies:
\begin{eqnarray}
\label{Bianchi}
\delta^{ab}(R_{0a0c;b}+R_{0acb;c}+ R_{0ab0;c})=0 \nonumber\\  \to  \delta^{ab}R_{0a0c;b}=0,
\end{eqnarray}
where in its second line, Eqs.\eqref{DR00} are used. Utilizing Eqs.~\eqref{DR00} and \eqref{Bianchi} results in:
\begin{eqnarray}
\label{RabcTraceless}
 \delta^{ab} {\,}^{n}\!{\cal R}_{abc} =0,
\end{eqnarray}
where $n \in\{1,\cdots,4\}$ and the Einstein summation over repeated indices is considered.

The solution of Eq.~\eqref{Psi3Eq} with the imposed boundary condition in Eq.~\eqref{BoundaryCondition3} is given by:  
\begin{eqnarray}
    \Psi^{(3)} = i Q^{(3)} \Psi^{(0)},
\end{eqnarray}
where:
\begin{eqnarray}
\label{Q3operator}
    Q^{(3)} &=& 
    -\frac{m}{6}  {\,}^{1}\!{\cal R}_{abc} x^a x^b x^c 
    + \frac{1}{2} {\,}^{2}\!{\cal R}_{abc} {x^a x^b p^c}\nonumber\\
    &-& \frac{1}{m} {\,}^{3}\!{\cal R}_{abc}{ x^a p^b p^c} 
    + \frac{1}{m^2}{\,}^{4}\!{\cal R}_{abc} p^a p^b p^c.
\end{eqnarray}
Note that, from Eq.~\eqref{RabcTraceless},  $Q^{(3)}$ is Hermitian, i.e., $Q^{(3)}=Q^{\dagger(3)}$. In other words, only the symmetric products of $x^a$ and $p^b$ contribute to $Q^{(3)}$.
The commutator appearing in the residual force then is given by: 
\begin{eqnarray}
\label{Q3Commutator}
     i[p_a, Q^{(3)}]
    \!&=&\! -\frac{m {\,}^{1}\!{\cal R}_{abc} x^b x^c}{2} 
    + \frac{1}{2}{\,}^{2}\!{\cal R}_{abc} {(x^b p^c+p^c x^b)}\nonumber\\
    &-& \frac{{\,}^{3}\!{\cal R}_{abc}  p^b p^c}{m},  
\end{eqnarray}
which can be employed in Eq.~\eqref{F3} to calculate the residual force at the cubic order's correction.

Let us first consider spherical modes. In the standard Hermite basis presented in Eq. \eqref{StandardHermitBasis}, only those with the same excitation numbers for all directions are spherical. So any initially spherical state can be written as,
\begin{eqnarray}
    |S^{(0)} \rangle = \sum_{n=0}^{\infty} c_n \ket{n_s},
\end{eqnarray}
where  $n_s=(n,n,n)$, and $c_n$ are expansion coefficients, and $\ket{n_s}$ is the corresponding Hermite mode. The residual force for a spherical state vanishes due to Eq.~\eqref{RabcTraceless}. So initially, spherical states do not experience any force at the cubic order in the transverse direction and keep moving along the central time-like geodesic.

Any non-spherical state, however, generally feels a force. To calculate the force for a general state, we notice that  Eqs.~\eqref{vanishingP} and \eqref{vanishingX} imply:
\begin{eqnarray}
\label{Eq152Aux}
 \langle \Psi^{(0)}|x^a x^b |\Psi^{(0)}\rangle &=& \langle \Psi^{(0)}|x^a p^b |\Psi^{(0)}\rangle\\&=&  \langle \Psi^{(0)}|p^a p^b |\Psi^{(0)}\rangle  = 0,~\forall a\neq b\nonumber.
\end{eqnarray}
In the following, we would like to calculate the force exerted on the standard Hermite mode $|\Psi^{(0)}\rangle=\ket{\vec{n}}$, where $\vec{n}=(n_1, n_2, n_3)$ while $n_a$ represents the mode's number in $x_a$'s direction. Thus, we first substitute Eq.~\eqref{Q3Commutator} in Eq.~\eqref{F3} and rewrite the force at the cubic order's correction by:
\begin{subequations}
\label{TotalForce}
\begin{eqnarray}
    F_a^{(3)}(\vec{n})= {}^{1}\!F_a^{(3)}+{}^{2}\!F_a^{(3)}+{}^{3}\!F_a^{(3)},
\end{eqnarray}
where 
\begin{eqnarray}
    {}^{1}\!F_a^{(3)}&=&-\frac{\tau_{\!_R}}{w_0^2}\partial_\tau\left({\,}^{1}\!{\cal R}_{abc}\sandwich{\vec{n}}{x^b x^c}{\vec{n}}\right), \\
    {}^{2}\!F_a^{(3)}&=&\frac{1}{2}\partial_\tau\left({\,}^{2}\!{\cal R}_{abc}\sandwich{\vec{n}}{(x^b p^c + p^c x^b)}{\vec{n}}\right),\\
    {}^{3}\!F_a^{(3)}&=&-\frac{w_0^2}{2\tau_{\!_R}}\partial_\tau\left({\,}^{3}\!{\cal R}_{abc}\sandwich{\Vec{n}}{p^b p^c}{\vec{n}}\right),
\end{eqnarray}
\end{subequations}
where $m$ in \eqref{Q3Commutator} is substituted with $\frac{2\tau_{_R}}{w_0^2}$. 
Employing Eq.~\eqref{Eq152Aux} and the exact form of $\ket{\vec{n}}$ yield:
\begin{eqnarray}
\sandwich{\vec{n}}{x^b x^c}{\vec{n}} &=& \frac{2n_{b}+1}{4} w^2 \delta^{bc},\\
\frac{1}{2}\sandwich{\vec{n}}{(x^b p^c+p^c x^b)}{\vec{n}}  &=&\frac{(2n_{b}+1)\tau}{2\tau_{\!_R}} \delta^{bc} ,\\
\sandwich{\vec{n}}{p^b p^c}{\vec{n}} &=& \frac{2n_{b}+1}{w_0^2}\delta^{bc} ,
\end{eqnarray}
where $\tau_{\!_R}$ is the Rayleigh proper time presented in Eq.~\eqref{RayleighTime}. 
The force exerted on $\ket{\vec{n}}$, then, is simplified to:
\begin{eqnarray}
\label{F3Fabc}
    F_{a}^{(3)} &=& F_{abc} n_{b} \delta^{bc},
\end{eqnarray}
where summation over $b$ and $c$ are understood, and:
\begin{equation}
\label{Fabc}
    F_{abc} = -\frac{\tau_{\!_R}^2+\tau^2}{2 \tau_{\!_R}} R_{0(a0b;c)}.
\end{equation}
We observe that only ${R}_{0(a0b;c)}$ contributes to the residual force at the cubic order's correction. The integration of ${R}_{0(a0b;c)}$ over the geodesic that is encoded in the ${\cal{R}}_3$ functions do not contribute to the residual force. The residual force at the cubic order's correction remains local because the force does not depend on the history of the wave function.  At the sub-sub-leading order, the square of $\Psi^{(2)}$ contributes to the equation of motion for $\Psi^{(4)}$. Since $\Psi^{(4)}$ depends on the history of the wave function by the ${\cal R}_2$ functions defined in Eq. \eqref{R2functions}, we expect that the residual force depends on the history of the wave function at the order of $\epsilon^4$. However, performing the calculation at order $\epsilon^4$ is beyond the scope of the current work.

\section{Schwarzschild spacetime geometry}
\label{section:schwarzchild}
The line element of the Schwarzschild black hole \cite{SchwarzschildBHL}, which also describes the spacetime geometry outside a spherically stationary  mass distribution with net mass $M$ in the standard spherical coordinates, is given by:
\begin{eqnarray}
 ds^2 = -\left(1-\frac{r_s}{r}\right) dt^2 + \frac{dr^2}{\left(1-\frac{r_s}{r}\right)} + r^2 d\Omega^2,
\end{eqnarray}
where $d\Omega^2 =d\theta^2+\sin^2\theta\, d\phi^2$ and $r_s= 2 G M$, $G$ is the Newton gravitational constant, and $r_s$ is called the Schwarzschild radius. We choose the unit of length such that $r_s=1$. 

Let us now consider geodesics in the Schwarzschild geometry. Due to the spherically symmetry, without loss of generality, we can choose the equatorial plane, i.e.:
\begin{eqnarray}
\label{thetadot}
    \theta&=& \frac{\pi}{2},\nonumber\\
    \dot{\theta}&=&0,
\end{eqnarray}
to describe any given geodesic at all times. The cyclic variables of $\phi$ and $t$ lead to invariant quantities:
\begin{eqnarray}
	\label{Eq6.5}
	\frac{\partial {\cal L}}{\partial \phi} = 0 &\to& r^2 \dot{\phi} = l \,,\\
	\label{Eq6.6}
	\frac{\partial {\cal L}}{\partial t} = 0 &\to& \left(1-\frac{1}{r}\right) \dot{t} = E\,.
\end{eqnarray}
\textcolor{black}{The Lagrangian being independent of $\tau$ implies that the Hamiltonian exhibits time invariance. Additionally, since the Lagrangian has a quadratic form, the Hamiltonian can be equated to the Lagrangian. Consequently, we can establish the following}:
\begin{equation}
	\label{Eq6.7}
	{\cal L} = -\left(1-\frac{1}{r}\right)\dot{t}^2+ \frac{\dot{r}^2}{1-\frac{1}{r}}  + r^2 \ \dot{\phi}^2 = -1\,.
\end{equation}
The non-zero components of the first covariant derivative of the Riemann tensor  evaluated on the geodesic ($\theta=\frac{\pi}{2}$) are given by:
\begin{eqnarray}
\label{DR}
    R_{t\theta t \theta;r} &=& R_{t\phi t \phi;r} = R_{\theta t r t;\theta} = R_{\phi t r t;\phi}= -\frac{3(r-1)}{2r^3},\nonumber\\
    R_{r \theta r \theta;r} &=& R_{r \phi r \phi;r} = \frac{3}{2r(r-1)},R_{trtr;r} = \frac{3}{r^4},\nonumber\\
    R_{\theta\phi\theta\phi;r} &=& 2R_{\phi\theta\phi r;\theta}=2 R_{\phi\theta \theta r;\phi} =-3.
\end{eqnarray}
\textcolor{black}{We assume that in the far past, the laboratory was located at asymptotic infinity and was moving towards a black hole with a finite non-zero angular momentum per unit mass. Therefore, the laboratory's time-like geodesic can be described by a finite parameter $l$ while $l$ is not equal to zero.} We further assume that the absolute value of the velocity of the lab at the asymptotic infinity is $v$, and the lab is radially falling into the black hole.  Equation \eqref{Eq6.5} then implies that the lab velocity is radial at the asymptotic infinity since $r \dot{\phi}$ vanishes at the asymptotic infinity. So it  holds that,
\begin{eqnarray}
\label{gamma}
    \dot{r}|_{r=\infty} &=&- \gamma v,~
	\dot{t}|_{r=\infty} = \gamma ,
\end{eqnarray}
where $\gamma$ is the boost factor: $\gamma=\frac{1}{\sqrt{1-v^2}}$. Equation \eqref{Eq6.6}, then, implies $E=\gamma$ and yields 
$	\dot{t} = \frac{\gamma r}{r-1}$. Equation~\eqref{Eq6.7} results:
\begin{equation}
\label{rdot}
    \dot{r}^2 = \gamma^2 - \left(1+\frac{l^2}{r^2}\right)\left(1-\frac{1}{r}\right).
\end{equation}
To calculate the components of the first derivative of the metric in the Fermi coordinates from \eqref{DR}, we use the tensorial property of the covariant derivative of the Riemann tensor. Assuming that the components of the covariant derivative of the Riemann tensor in the general coordinates $x^\mu$ is given by $R_{\mu\nu\lambda\eta;\gamma}$, the components of the covariant derivative of the Riemann tensor in the coordinates $y^{\mu'}$ is given by:
\textcolor{black}{
\begin{equation}
\label{eq163}
    R_{\mu'\nu'\lambda'\eta';\gamma'}= 
    \frac{\partial x^\mu}{\partial y^{\mu'}}
    \frac{\partial x^\nu}{\partial y^{\nu'}}
    \frac{\partial x^\lambda}{\partial y^{\lambda'}}
    \frac{\partial x^\eta}{\partial y^{\eta'}}
    \frac{\partial x^\gamma}{\partial y^{\gamma'}}
    R_{\mu\nu\lambda\eta;\gamma},
\end{equation}
where $R_{\mu'\nu'\lambda'\eta';\gamma'}$} represents the components of the covariant derivative of the Riemann tensor in $y^{\mu'}$ coordinates. 
The components of the covariant derivative of the Riemann tensor in the standard spherical coordinates are given in \eqref{DR}. We  utilize  \cite{Exirifard:2021cav}) to write the partial derivative of the standard spherical coordinates with respect to the Fermi coordinates: 
\begin{eqnarray}
\label{FermiStandardRelations}
    \frac{\partial r}{\partial x^0}&=&\! -\gamma \tanh \alpha \cos \beta,\frac{\partial r}{\partial x^1}=\!\gamma \cos \beta, \frac{\partial r}{\partial x^2}=-\frac{\gamma\sin \beta}{\cosh \alpha},\nonumber\\
    \frac{\partial \phi}{\partial x^0}&=&\! -\frac{\sinh \alpha  \sin \beta}{r}, \frac{\partial \phi}{\partial x^1}=\!\frac{\cosh \alpha \sin \beta}{r},\frac{\partial \phi}{\partial x^2}=\!\frac{\cos \beta}{r},\nonumber\\
    \frac{\partial t}{\partial x^0}&=& \frac{\cosh^2\alpha}{\gamma},~\frac{\partial t}{\partial x^1}=-\frac{\sinh 2 \alpha}{2\gamma},\frac{\partial \theta}{\partial x^3}= \frac{1}{r}.
\end{eqnarray}
Here, $\alpha$ and $\beta$ are defined by:
\begin{eqnarray}
	\cosh \alpha &=& \gamma \sqrt{\frac{r}{r-1}}, 
~	\sinh \alpha = \sqrt{\cosh^2 \alpha-1},\nonumber\\
	\label{Eq6.22b}	
	\sin \beta &=&\frac{l }{r \sinh \alpha},~ \cos \beta = \sqrt{1-\sin^2 \beta}.
\end{eqnarray}	
To succinctly express the non-zero components of $R_{0(a0b;c)}$ which appear in the force and in Eq.~\eqref{Gabc}, we first define variable $\chi$ by:
\begin{eqnarray}
    \chi = \frac{\sin \beta}{(r-1)\cosh \alpha}=\frac{l}{r}
    \left(\gamma^2  -1 +\frac{1}{r}\right)^{-\frac{1}{2}},
\end{eqnarray}
and notice:
\begin{eqnarray}
    R_{0(101;1)}+R_{0(202:1)}+R_{0(303;1)}&=&0,\nonumber\\
  R_{0(101;2)}+R_{0(202:2)}+R_{0(303;2)}&=&0.
\end{eqnarray}
The independent non-zero components of $R_{0(b0b;c)}$ then are given by:
\begin{eqnarray}
\label{R0a0bcSchw}
    R_{0(101;1)}&=&\frac{3 \gamma  \cos\beta \left(5 \cos^2\!\beta -3\right)}{2 r^{4}},\nonumber\\
    R_{0(303;1)} &=& -\frac{ \gamma \cos \beta \left(5l^{2}+3 r^{2}\right) }{2 r^{6}},\nonumber\\
    R_{0(303;2)} &=& \frac{3 \gamma^3 \chi}{2r^3}\left(4- \frac{3(r-1)}{r\gamma^2}\right),\nonumber\\
    R_{0(202;2)} &=&\frac{3\gamma^3\chi}{2r^3} \left( 5 \cos^{2}\!\beta  -3 +\frac{r-1}{r\gamma^2}\right). 
\end{eqnarray}
We have directly calculated and checked that all components of $R_{0(a0b;c)}$ remain finite on and outside the event horizon. Note that $R_{0(a0b;3)}=0$; therefore, no residual force exists in the direction of $x^3$. So at this level, the plane of the trajectory of the localised quantum wave remains intact.

\subsection{Radially falling localised wave function}
\label{section:radialSch}
 \begin{figure}[t]
 	\centering
 	\includegraphics[width=.90\linewidth]{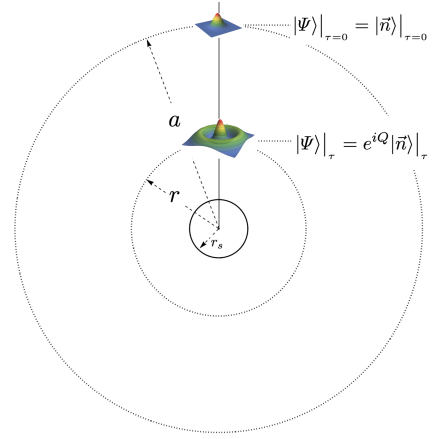}
 	\caption{A localised wave function at $r=a$ and $\tau=0$, which coincides with a Hermite-Gaussian mode $\ket{\vec{n}}$ with initial width of $w_0$, freely falls into the Schwarzschild black hole on a radial geodesic. As the wave function freely falls into the black hole, it gets distorted due to its interaction with the Riemann tensor around the geodesic. The quantum operator $Q$ gives the distortion, which has a systematic expansion series as presented in Eq. \eqref{QepsilonExpansion}. The leading and sub-leading corrections to $Q$ are presented in Eqs. \eqref{QOperator} and \eqref{Q3operator}. Applying Newton's {\it Lex Secunda} on the distorted wave function yields a residual force as shown in Eq. \eqref{residaulForce}. The residual force is radial for the radial fall into the Schwarzschild black hole. It vanishes at the leading order and its value at sub-leading level is computed in Eq. \eqref{residualForceRadial}.}
 	\label{Fig:RadialFalling}
 \end{figure}
\textcolor{black}{
In order to consistently describe the physics of a localised wave function/wavepacket in the presence of a Schwarzschild black hole, we consider that: 
\begin{enumerate}
    \item In the rest frame of the black hole, the energy associated to the wave function is much smaller than the mass of black hole.
    \item The width of the wavepacket is always smaller than the curvature of the spacetime geometry. 
    \item The central geodesic of the rest frame of the wave function is described by a time-like geodesic with parameters of $l$ and $\gamma$ where $l$ is defined in eq. \eqref{Eq6.5} and $\gamma$ is the boost factor with respect to the black hole at the asymptotic infinity. 
    \item The initial width of the wavepacket at distance $a$ from the black hole is given by $w_0$.
\end{enumerate} 
In this section, we assume that the central geodesic of the wavepacket is radially falling and is associated to $\gamma=1$ in \eqref{gamma}. The central radial geodesic, therefore, is characterised by:
\begin{eqnarray}
\label{radialgeodesic}
l&=&\beta=0,\nonumber\\
\gamma &=&1.
\end{eqnarray}
The Fermi coordinates then coincide to $\hat{x}^1=\hat{r}$, $\hat{x}^2=\hat{\phi}$ and $\hat{x}^3=\hat{\theta}$ \cite{Exirifard:2021cav}.  Equation \eqref{rdot} can be solved to find the proper time that the wavepacket needs to reach the Schwarzschild radius/event horizon: 
\begin{eqnarray}\label{tauasfr}
    \tau_s = \frac{2}{3}(a^{\frac{3}{2}}-1).
\end{eqnarray}
We assume that $a$ is much larger than $1$, so the width of the wavepacket at the event horizon can be approximate by:$ w_s=\frac{4 a^{3/2}}{3 m w_0}-\frac{4}{3 m w_0}+O(a^{-3/2}).$  The consistency of our approach requires  that the width of the wave packet  remains smaller than the curvature of the spacetime geometry. We utilize  the Kretschmann invariant to estimate the curvature of the spacetime geometry: 
$R_{\mu\nu\alpha\beta} R^{\mu\nu\alpha\beta}= \frac{12}{r^6}$.
We require the width of the wavepacket on the event horizon to be much smaller than the square root of the inverse of the Kretschmann invariant: 
\begin{equation}
\label{ConsistencyCondition}
    \left(\frac{a}{r_s}\right)^{\frac{3}{2}} \ll \frac{\sqrt{3}}{8}  \frac{w_0}{\lambda_c},
\end{equation}
\textcolor{black}{Notice that symbol $\lambda_c$ represents the Compton wavelength associated with a particle of mass $m$. It is defined as $\lambda_c=\frac{\hbar c}{m}$, where $\hbar$ is the reduced Planck constant and $c$ is the speed of light. In addition, the Schwarzschild radius $r_s$ is explicitly shown too.}  If we apply \eqref{ConsistencyCondition} to the wave packets of the protons in the boundary of the Milky Way, we see that only the ones  with an initial width  larger than $20$ meters may fall into Sagittarius $A^*$. 
\\
We assume that \eqref{ConsistencyCondition} is satisfied.}  Furthermore, for sake of simplicity we consider the initial standard Hermite mode $\ket{\vec{n}}$ with an initial width of $w_0$:
\begin{eqnarray}
\ket{\Psi}\big{|}_{\tau=0} = \ket{\vec{n}}.
\end{eqnarray}
We utilise \eqref{PsiQ} to represent the wave function at $\tau>0$:
\begin{eqnarray}
\ket{\Psi} = \exp(i Q) \ket{\vec{n}},
\end{eqnarray}
where $Q$ encodes the evolution of the wave function due to its interaction with the Riemann tensor, the $\epsilon$ expansion series for $Q$ is provided in Eq.~\eqref{QepsilonExpansion}. Fig. \ref{Fig:RadialFalling} depicts the wavepacket at $r=a$, and a later time at a general $r$. We are interested in calculating the residual force, which is presented in Eqs.~\eqref{F3Fabc} and \eqref{Fabc}. 
The non-zero components of $R_{0(a0b;c)}$ provided in Eq.~\eqref{R0a0bcSchw} for Eq.~\eqref{radialgeodesic} are simplified to:
\begin{eqnarray}
     R_{0(101;1)}=-2R_{0(202;1)}=-2R_{0(303;1)}=\frac{3}{r^4},
\end{eqnarray}
which implies that non-zero coefficients of $F_{abc}$ introduced in Eq. \eqref{Fabc} hold:
\begin{eqnarray}
    F_{221}=F_{331}=-\frac{1}{2} F_{111}.
\end{eqnarray}
This means that the residual force exists only in the direction of $x^1$ that coincides with the $r$'s direction.  Utilizing \eqref{F3Fabc}, one can calculate the force exerted on the state $\ket{\vec{n}}$:
\begin{eqnarray}
\label{residualForceRadial}
  F^{(3)}_1 &=& F_{111} \left(n_1-\frac{n_2+n_3}{2}\right),
\end{eqnarray}
where
\begin{eqnarray}
  F_{111} &=& -\frac{3(\tau_{\!_R}^2+\tau^2)}{2 \tau_{\!_R} r^4}.
\end{eqnarray}
$\tau$ is a function of $r$, as is explicitly expressed in Eq.~\eqref{tauasfr}. The divergence at $r=0$ is due to the singularity of the Schwarzschild metric at $r=0$.  It can also be shown that when the width of the wave function outside the event horizon remains much smaller than the Schwarzschild radius,  the impulse of the residual force does not generate an escape velocity for the in-falling wave functions, and the radially in-falling wave function falls into the event horizon.

\subsection{Slight deflection by a black hole}
\label{section:deflection}
 \begin{figure}[t]
 	\centering
 	\includegraphics[width=.90\linewidth]{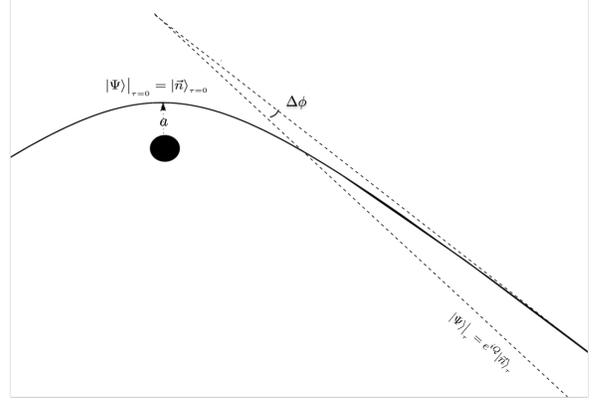}
 	\caption{A particle in mode of $\ket{\vec{n}}$ with initial width of $w_0$ is produced at distance $r=a$ and $\tau=0$ in the Schwarzschild geometry. The classical trajectory of the particle is given by a time-like geodesic along where the speed of the particle at the asymptotic infinity is $v$ with the boost factor of $\gamma=\frac{1}{\sqrt{1-v^2}}$, and $a$ is the minimum distance of the particle to the black hole on the trajectory. Due to the interaction of the wave function with the Riemann tensor around the time-like geodesic, the asymptotic trajectory of the mode $\vec{n}$ makes the angle $\Delta\phi$ with the classical trajectory. $\Delta\phi$ is calculated in Eq. \eqref{extradeflection}.}
 	\label{Fig:Scatering}
 \end{figure}
We assume that the wave function's trajectory (the central geodesic) always remains far away from the black hole. So we can assume $1\ll r$. We, therefore, approximate Eq. \eqref{rdot} to:
\begin{eqnarray}
\label{rdotapprox}
\dot{r}=\pm \sqrt{\gamma^2 - \left(1+\frac{l^2}{r^2}\right)}.
\end{eqnarray}
which vanishes at the minimum distance of the wave function to the black hole. So the minimum distance between the trajectory of central geodesic to the black hole is given by,
\begin{equation}
	\label{adistanceReferee1}
    a = \frac{l}{\sqrt{\gamma^2-1}}.
\end{equation}
We consider that the wave function is localised around $r=a$ at $\tau=0$ and moves away from the black hole, and coincides with the Hermite mode $\ket{n}$ with the initial width of $w_0$. So, Eq. \eqref{rdotapprox} results in:
\begin{eqnarray}
\dot{r} &= &\left\{
\begin{array}{l c r}
-\sqrt{\gamma^2 +\frac{l^2}{r^2}}   &,  & \tau \leq 0  \\
 \sqrt{\gamma^2 +\frac{l^2}{r^2}} &,    & 0 \leq \tau
\end{array}
\right.
\end{eqnarray}
So it yields:
\begin{eqnarray}
\tau^2 &= & \sqrt{\frac{r^2 - a^2}{\gamma^2 -1}}.
\end{eqnarray}
The large $r$ limit of Eq. \eqref{R0a0bcSchw} holds:
\begin{eqnarray}
R_{0(101;2)}&=& \frac{a \left(-9 \gamma ^2+3\right)}{r^5}, \\
R_{0(202;2)} &=&\frac{3 a \left(2 \gamma ^2+1\right)}{2 r^5}, \\
R_{0(303;2)} &=&\frac{3 a \left(4 \gamma ^2-3\right)}{2 r^5}.
\end{eqnarray}
To succinctly express the residual force, consider:
\begin{eqnarray}
q&=&-3 n_1+n_2+2 n_3,\\ 
p&=&2 n_1+n_2-3 n_3.
\end{eqnarray}
Then, the second component of the residual force given in Eq. \eqref{F3Fabc} simplifies to:
\begin{eqnarray}
F^{(3)}_2=\frac{3 a}{4 r^5 \tau_{_R}}(2 q\gamma^2 + p ) \left(\tau ^2+\tau_{_R}^2\right).
\end{eqnarray}
The momentum that the residual force generates is given by its impulse:
\begin{eqnarray}
\label{eq193}
\langle p_2\rangle&=&\int_0^{\infty} d\tau F^{(3)}_2,\nonumber\\
&=&-\frac{\left(a^2+2 \left(\gamma ^2-1\right) \tau_{_R}^2\right) \left(p+2 \gamma ^2 q\right)}{2 a^3 \left(\gamma ^2-1\right)^{3/2} \tau_{_R}}.
\end{eqnarray}
where $\infty$ in the boundary of the integration is due to approximating  $a\gg1$.  The residual force causes an extra deflection in the scattering of the wave function that can be approximated by: 
\begin{eqnarray}
\label{extradeflection}
\Delta\phi = \frac{\langle p_2\rangle}{\gamma m v}.
\end{eqnarray}
We observe that the deflection that is caused by the residual force is a function of $n_1$, $n_2$ and $n_3$. \textcolor{black}{Notice that $v$ is the velocity of the rest frame (at the asymptotic infinity) with respect to the black hole. Therefore, $v$ can be arbitrary large.} One may wish to consider the large $\gamma$ limit of \eqref{extradeflection} in order to possibly gain some insights on the deflection of a spatially structured photon by a black hole:
\begin{eqnarray}
\label{predicted_effect}
\lim_{\gamma\to \infty} \Delta\phi = -\frac{q w_0^2}{a^3}.
\end{eqnarray}
Considering that the light deflection in Einstein theory of gravity is proportional to the inverse of $a$, besides the fact that $w_0$ should be much smaller than $a$, it casts a shadow of despair on the prospect of measuring \eqref{predicted_effect} by the available technologies. The precision of the E\"otv\"os experiment near Earth is also not sensitive to the predicted displacement due to different internal quantum structures  \cite{Emelyanov:2021auc}.
\section{Conclusions}
\label{section:conclusions}
Propagation of a localised wave function of \textcolor{black}{a massive particle} been studied in the flat and curved spacetime geometry. The notion of a rest frame is extended for a localised wave function in section \ref{section:restframe}. The low energy modes in the rest frame of the wave function have been studied, and complete orthogonal basis of the standard Hermite-Gaussian modes has been adapted in section \ref{section:rfeq}.  We have utilised the Fermi coordinates adapted to the time-like geodesic of the rest frame and have calculated the leading and sub-leading general relativistic corrections to a general, localised wave function in section \ref{Section:curvedgeometry}.  It has been shown that as a localised wave function freely propagates in a curved spacetime geometry, it interacts with the curvature of the spacetime geometry around the geodesic. This interaction leads to the distortion operator $Q$ defined in  \eqref{PsiQ}. The quadratic and cubic order corrections to the distortion operator have been calculated in  \eqref{QOperator} and \eqref{Q3operator}. It has been shown that the distortion operator is a function of the Riemann tensor, and its covariant derivative is evaluated and integrated over the time-like geodesic of the rest frame.

Section \ref{Section:residualforce} has employed Newton's {\it Lex Secunda} in order to define the residual net force acting on the wave function $\Psi$ as the time rate of the change of its average linear momentum. It has been shown that the residual net force vanishes at the level of quadratic corrections. However, it does not vanish at the order of the cubic corrections for a general wave function, as shown in  \ref{secion:CubicCorrections}. It has been proved that spherical symmetric wave functions follow the time-like geodesic of the rest frame. However, non-spherically symmetric wave functions have been shown to experience a net residual force and deviate from the time-like geodesic of the rest frame. 

The example of the Schwarzschild spacetime geometry has been studied in section \ref{section:schwarzchild}. The residual force has been computed for radial fall into the black hole \ref{section:radialSch} and a slight deflection from the black hole \ref{section:deflection}. It has been shown that the residual force does not generate an escape velocity for the in-falling particle. However, the residual force for the scattered wave functions leads to a mode-dependent deflection angle where the dependency is in \eqref{extradeflection}, as depicted in Fig. \ref{Fig:Scatering}.
Let us highlight that the dependency to the mode numbers resembles angular deviations in optics where different localised modes \cite{ObservedGoos} 
are observed to move along different paths in the non-homogeneous medium. Here, we are reporting the displacement of a localised massive scalar from its expected time-like geodesic. The computed dependency of the deflection in the Solar system turns to be very small to be measured by available technologies. Nonetheless, the finding that trajectories of localised wave functions deviate from geodesics can be counted as evidence supporting the idea that gravity should not be treated as a geometrical force in the quantum realm.  

\acknowledgments 
This work was supported by the High Throughput and Secure Networks Challenge Program at the National Research Council of Canada, the Canada Research Chairs (CRC) and Canada First Research Excellence Fund (CFREF) Program, and the Joint Centre for Extreme Photonics (JCEP). We thank Alicia Sit for proofreading the paper.

\end{document}